\documentclass[aip,reprint,onecolumn,english,nofootinbib,letterpaper,nobalancelastpage]{revtex4}
\usepackage{lmodern}

\usepackage[T1]{fontenc}
\usepackage[latin9]{inputenc}
\usepackage[letterpaper]{geometry}
\usepackage{amsmath}
\usepackage{amssymb}
\usepackage{bbm}
\usepackage{babel}
\usepackage{graphicx}

\usepackage{latexsym}
\usepackage{mathptm}

\newcommand{\f}{\begin{equation}}
\newcommand{\ff}{\end{equation}}

\begin{document}
\title{Particle-dependent deformations of Lorentz symmetry}

\author{Giovanni Amelino-Camelia}
%\email{Giovanni.Amelino-Camelia@roma1.infn.it}
\affiliation{\footnotesize{Dipartimento di Fisica, Universit\`a ``La Sapienza", P.le~Moro~2,~Roma,~EU}}
\affiliation{\footnotesize{INFN, Sez.~Roma1, P.le Moro 2, 00185 Roma, EU}}

\begin{abstract}
I here investigate what is arguably the most significant residual challenge
for the proposal of phenomenologically viable ``DSR deformations"
of relativistic kinematics, which concerns the description of
composite particles, such as atoms.
In some approaches to the formalization of possible scenarios for DSR-deformation
of Lorentz symmetry it emerges that composite particles should have relativistic
properties different from the ones of their constituent ``fundamental particles",
but these previous results
provided no clue as to how the mismatch of relativistic properties could be consistently
implemented. I show that
it is possible to implement a fully consistent DSR-relativistic
description of kinematics endowing different types of particles
with suitably different deformed-Lorentz-symmetry properties.
I also contemplate the possibility that some types of particles (or macroscopic bodies)
behave according to completely undeformed special relativity,
which in particular might apply to the DSR description of the macroscopic bodies
 that constitute measuring devices (``observers").
The formalization is also applicable to cases where different fundamental
particles have different relativistic properties, leading to a type of phenomenology
which I illustrate by considering possible applications  to the ongoing analyses of
the ``Lorentz-symmetry anomaly" that was recently tentatively reported by the OPERA collaboration.
Some of the new elements here introduced in the formulation of relativistic kinematics
appear to also provide the starting point for the development of a correspondingly
novel mathematical formulation of spacetime-symmetry algebras.
 \end{abstract}

\maketitle

\vskip -0.95cm

\tableofcontents

\newpage

\section{Introduction}
One of the most active areas of quantum-gravity research over the last decade
concerns the fate of Lorentz symmetry in the quantum-gravity realm\footnote{The relevant aspect of the quantum-gravity
 problem is the ``quasi-Minkowski limit" of quantum gravity, the limit where quantum gravity should
 reproduce in first approximation
 (describing small modifications of) particle physics and its special-relativistic properties.}.
It is turning out to be particularly useful
to divide all such studies into three categories: (i) cases where Lorentz/Poincar\'e
symmetry remains unaffected; (ii) cases where there are departures from
classical Lorentz/Poincar\'e symmetry  and they are such that a preferred-frame picture arises;
(iii) cases where there are departures from
classical Lorentz/Poincar\'e symmetry but the relativity of inertial
frames is preserved. This third option, which was proposed in Refs.~\cite{dsr1Edsr2},
is the one that challenges us more significantly for what concerns formalization.
The case of ``broken Lorentz symmetry" (with a preferred frame)
is technically not much more challenging
that a standard special-relativistic case,
since it
 allows formalizations that are rather familiar, already relevant for example
  in the analysis of conventional propagation of light in certain material
media (which indeed provide a preferred frame for the analysis of those physical contexts).
Instead formalizing ``deformations of Lorentz symmetry", in the sense of the ``DSR"
(``doubly-special", or ``deformed-special", relativity)
proposal of Refs.~\cite{dsr1Edsr2}, requires us to find ways of introducing departures
from Lorentz/Poincar\'e
symmetry while preserving the delicate balance that can assure the relativity of inertial frames.
This DSR proposal focuses on the possibility of relativistic theories with
two characterizing invariant scales, introducing a length/inverse-momentum scale $\ell$
with relativistic properties analogous to the familiar ones of the speed-of-light scale $c$.
From a quantum-gravity perspective it would then be natural~\cite{dsr1Edsr2}
to assume that the new relativistic-invariant scale $\ell$
be roughly of the order of the inverse of the Planck scale (the ``Planck length").

At this point there is a rich literature on DSR-deformations of Lorentz symmetry,
with several encouraging results (see, {\it e.g.},
Refs.~\cite{dsr1Edsr2,jurekdsr1,dsrPOLAND2001,leeDSRprd,jurekDSR2,leeDSRrainbow,gacdsrrev2010}
and references therein).
Most of these results concern DSR-relativistic formulations of
the possibility of introducing relativistically some deformed on-shell relations
and some associated deformations of the laws of composition of momenta.
The  DSR proposal was put forward~\cite{dsr1Edsr2} as a conceptual path for pursuing
a broader class of scenarios of interest for fundamental physics, and in particular for
quantum-gravity research, including the possibility of introducing the second
observer-independent scale primitively in spacetime structure or primitively at the
level of the (deformed) de Broglie relation between wavelength and momentum.
However, the bulk of the relevant preliminary results from quantum-gravity research
concern departures from the special-relativistic on-shell relation, and this in turn became
the main focus of DSR research.

I here investigate issues relevant for one of the most significant residual
open issue for such studies of DSR-relativistic deformations
of on-shell (and momentum-conservation) relations,
which concerns the description of
composite particles, such as atoms.
In some of the most studied attempts of formulating DSR-relativistic theories
it emerges  that ``DSR-composite particles" should have relativistic
properties different from those of their constituents.
The simplest way to see that this might be the case is to consider a bunch
of $N$ ultrarelativistic particles all propagating along the $1$ direction
and each governed, say, by
\begin{equation}
p_0 \simeq p_1 +\frac{m^2}{2p_1} - \ell p_1^2
\label{intro1}
\end{equation}
We can then introduce some candidates for ``total spatial momentum" and ``total energy",
given by ${\cal P}_0 = N p_0$ and ${\cal P}_1 = N p_1$, and observe that
the validity of (\ref{intro1}) for each of the $N$ particles implies
\begin{equation}
{\cal P}_0 \simeq {\cal P}_1 +\frac{\mu^2}{2{\cal P}_1} - \frac{\ell {\cal P}_1^2}{N}~,
\label{intro2}
\end{equation}
where $\mu = N m$ is the rest energy of the $N$-particle system. The suppression by $1/N$
of the last term of this Eq.~(\ref{intro2}) illustrates the issue: the nonlinearity
of the DSR laws has nontrivial consequences for particle composites.\\
 Much more than this
simple-minded argument supports the concern that composite particles should have
relativistic properties which are different from those of their constituents,
and in particular the effects of the deformation should be more weakly felt by composites.
I shall not review here these more sophisticated arguments, for which I refer my readers
to Refs.~\cite{dsrPOLAND2001,joaoMACRO}
and, most notably, Ref.~\cite{soccerball}.
Let me stress however that these technical arguments, based on the nonlinearity of the laws
and the way it can affect the description of composites, also makes sense physically:
while an on-shell relation of type (\ref{intro1}) is certainly plausible, at least if $\ell$ is indeed of the order of the inverse of
the Planck scale, for
microscopic particles, the same on-shell relation, even taking $\ell$ as the inverse of
the Planck scale, is unacceptable for macroscopic bodies composed of very many
micro-particles. The Planck-scale is huge by the standards of elementary particles
but is actually a small scale ($\sim 10^{-5} grams$) for macroscopic bodies, and as
a result, unless there is a suppression of the type shown in
 Eq.~(\ref{intro2}), the DSR description of macroscopic bodies could be disastrous.

 So there is technical evidence of the fact that composite particles
should have DSR-relativistic
properties different from those of their constituents,
with weaker deformation effects, and this is much welcome from the
point of view of reproducing the observed properties of macroscopic bodies.
But this encouraging correspondence between features for composite (particles and) bodies
found on the theory side and our desiderata for the phenomenology of macroscopic bodies
has also provided a formidable challenge for DSR research: if composite particles (and
macroscopic bodies) have relativistic properties which are different from the ones
of their constituents then these DSR-relativistic theories should be theories that
do not prescribe "universal" laws of kinematics but rather particle-dependent ones!!\\
Is that even possible?\\
{\underline{Can a theory be fully relativistic and yet attribute different laws
of kinematics to different particles?}}\\
These questions have remained so far unanswered.\\
I shall here show that {\underline{the correct answer is yes}:
there are logically consistent DSR-relativistic
theories in which different particles (possibly ``elementary" and ``composite" particles)
are governed by different laws of kinematics.

After a brief reminder, in the next Sec.~\ref{dsrgeneral}, of the basic logical structure
of DSR-relativistic theories, I set the stage for my analysis, in Sec.~\ref{kappauno},
by reviewing in some detail the relativistic kinematics of a much studied DSR framework.
This is of course still a standard ``universal" DSR framework, but a rather sophisticated
possibility. My attitude here is not one of establishing the validity of general theorems,
so the strength of my results is primarily exhibited in terms of the fact that I can take
as starting point a rather sophisticated ``universal" DSR framework. The fact that I am able
to generalize
such a framework to a ``non-universal" version (with particle-type-dependent effects)
suggests that such generalizations might be even easier when taking as starting
point simpler ``universal" DSR scenarios.

A first group of new results is reported in Sec.~\ref{hopfliesec}, where I show that
one can combine in the same relativistic theory particles with standard special-relativistic
properties and particles with DSR-deformed relativistic properties.
The key ingredient of this result is a ``mixing composition law" suitable for writing
a DSR-covariant law of conservation of momentum for processes involving different particles
with different relativistic properties, and such that the covariance is assured by
 a suitably adapted action of boost generators on multiparticle systems.

I then show, in Sec.~\ref{hopfhopfsec}, that
 one can combine in the same relativistic theory two species of particles; one
  with some given DSR-deformed relativistic properties and the other with some other
DSR-deformed relativistic properties. I verify that this is possible at least in cases
such that the DSR deformation has the same formalization for the two types of particles,
but with different magnitude.
I introduce for this purpose a further
generalization of the ``mixing composition law" and of the laws of
action of boost generators on multiparticle systems.

The new formulation of relativistic kinematics introduced
in Secs.~\ref{hopfliesec} and~\ref{hopfhopfsec}
is most simply viewed from the perspective of applications to different types
of particles (some ``elementary" and some ``composite"),
but  I also explore, in Sec.~\ref{minibasketsec}, the possibility
that the laws characterized by
weaker DSR deformation apply to a macroscopic body. I find preliminary
encouragement for such a possible application.

Then, in Sec.~\ref{operasec}, I consider the possibility of applications of the formulation
of relativistic kinematics here introduced
to the case of different types of elementary particles, establishing a few first points
relevant for the phenomenology. As a way to illustrate more vividly the content of
this possible application I
use as ``conceptual laboratory"
the neutrino-superluminality-anomaly recently {\underline{tentatively}} reported by the
OPERA collaboration~\cite{opera}.

In the brief Sec.~\ref{mixingcoprodsec} I comment on the type of spacetime-symmetry
algebra which could provide the formal/mathematical counterpart for the
version of relativistic kinematics I here introduce.
In this respect perhaps most notably I argue that a suitable generalization
of the Hopf-algebra notion of co-product, something of the sort of a ``mixing co-product",
could be inspired by the ``mixing composition laws" I here introduced.

I work throughout at leading order in the deformation scale $\ell$ (with {\underline{$\ell$ that
can be both positive and negative}}, in the sense than both scenarios
with $\ell/|\ell| =1$ and scenarios with $\ell/|\ell| =-1$ are admissible).
This keeps formulas at reasonably manageable level, sufficiently characterizes the
new concepts, and would be fully sufficient for phenomenology
if indeed the deformation scale  is roughly of the order
of the huge Planck scale (in which case a leading-order analysis should be all we
need for comparison to data we could realistically imagine to gather
over the next few decades).\\
However, in Sec.~\ref{allordersec}, I do offer a small aside contemplating
possible generalizations of my results to ``all-order analyses".

Some speculations about possible future developments are offered in the brief closing
Sec.~\ref{closingsec}.

I mostly focus on 1+1-dimensional cases, where all conceptual issues
here relevant are already present and can be exposed more simply. Therefore my momenta
will often have two components, $\{p_0,p_1 \}$, and when I briefly switch to consider cases
with more dimensions I will use the notation $\{p_0,p_j \}$.

\section{DSR-deformations of Lorentz symmetry}\label{dsrgeneral}
Before proceeding with the main part of the analysis, let me pause briefly, in this section,
for summarizing the main points originally made in Ref.~\cite{dsr1Edsr2}
concerning the consistency requirements
that the relativity of inertial frames imposes on the relationship between
the form of the on-shell(/dispersion) relation and the form
of laws of energy-momentum conservation.\\
This is one of the most used DSR concepts,
and plays a pivotal role in the analysis I report in the following sections.

This consistency between on-shell relation and laws of momentum conservation
that follows from insisting on the relativity of inertial frames is also
rather significant from the perspective of studies of the quantum-gravity problem,
where in some cases one finds ``preliminary theoretical evidence" of modifications
of the on-shell relation but usually not accompanied so far by any information
on whether or not there should also be
 modifications of the law of conservation of momentum.
Indeed the idea of DSR-deformed Lorentz transformations was put forward~\cite{dsr1Edsr2} as a possible
description of certain {\underline{preliminary}}
theory results suggesting that there {\underline{might}} be violations
of some special-relativistic laws in certain approaches to the quantum-gravity problem,
most notably the ones based on spacetime noncommutativity and loop quantum gravity.
The part of the quantum-gravity community interested in those results was interpreting them
as a manifestation of a full breakdown of Lorentz symmetry, with the emergence of
a preferred class of observers (an ``ether"). But it was argued in Ref.~\cite{dsr1Edsr2}
that departures from Special Relativity governed by a high-energy/short-distance scale
may well be compatible with the Relativity Principle, the principle of relativity
of inertial observers, at the cost of allowing some consistent modifications
of the Poincar\'e transformations, and particularly of the Lorentz-boost transformations.
And it was already observed in Ref.~\cite{dsr1Edsr2} that this in turn would 
require corresponding modifications of the laws of momentum conservation.

As mentioned above, the  DSR proposal could provide~\cite{dsr1Edsr2}
a conceptual path for pursuing
a broader class of scenarios of interest for fundamental physics, and in particular for
quantum-gravity research, including the possibility of introducing the second
observer-independent scale primitively in spacetime structure or primitively at the
level of the (deformed) de Broglie relation between wavelength and momentum.
However, the bulk of the preliminary results from quantum-gravity research
concern departures from the special-relativistic on-shell relation, and this in turn became
the main focus of DSR research.

So let me consider a generic on-shell relation of the type
\begin{equation}
m^2 = p_0^2 - {\bf p}^2 + \Delta(E,{\bf p};\ell)
\label{dsr1gen}
\end{equation}
where $\Delta$ is the deformation and $\ell$ is the deformation scale.\\
Evidently when $\Delta \neq 0$ such an on-shell relation (\ref{dsr1gen})
is not Lorentz invariant. If we insist on this law and on
the validity of classical (undeformed) Lorentz transformations between inertial
observers we clearly end up with a preferred-frame picture, and the Principle
of Relativity of inertial frames must be abandoned: the scale $\ell$ cannot
be observer independent, and actually the whole form of (\ref{dsr1gen}) is subject
to vary from one class of inertial observers to another.\\
From the alternative DSR perspective one would have to enforce
the relativistic invariance of laws such as (\ref{dsr1gen}), preserving the relativity
of inertial frames, at the cost of modifying the action of boosts on momenta.
Then in such theories both the velocity scale $c$ (here mute only because of the
choice of dimensions) and the length/inverse-momentum scale $\ell$ play the
same role~\cite{dsr1Edsr2}
of invariant scales of the relativistic theory which govern the form of boost
transformations. \\
Several examples of boost deformations adapted in the DSR sense to modified on-shell
relations have been analyzed in some detail
(see {\it e.g.} Refs.~\cite{dsr1Edsr2,jurekdsr1,dsrPOLAND2001,leeDSRprd,jurekDSR2,leeDSRrainbow,gacdsrrev2010}
and references therein).
Clearly these DSR-deformed boosts ${\cal N}_j$ must be such that
\begin{equation}
[{\cal N}_j, p_0^2 - {\bf p}^2 + \Delta(E,{\bf p};M_*)] = 0~.
\label{boostsgen}
\end{equation}
This requirement (\ref{boostsgen}) of DSR-relativity
is completely analogous to the corresponding ones of Galilean Relativity and Special Relativity:
of course in all these cases
the on-shell relation is boost invariant (but respectively under Galilean boosts,
Lorentz boosts, and DSR-deformed Lorentz boosts); for Special Relativity the action of boosts
evidently must depend on the speed scale $c$ and boosts must act non-linearly on velocities
(since they must enforce observer-independence of $c$-dependent laws), and for DSR relativity
the action of boosts
evidently must depend on both the scale $c$ and the scale $\ell$, with boosts acting non-linearly
both on velocities and momenta, since it must enforce observer-independence of $c$-dependent and $\ell$-dependent laws.

Actually much of the logical structure of the
conjectured transition from Special Relativity to a DSR theory
can be understood in analogy with the transition from Galilean Relativity to Special relativity.
Famously, as the Maxwell formulation of electromagnetism,
with an observer-independent speed scale ``$c$", gained more and more
experimental support (among which one should count the Michelson-Morley results)
it became clear that Galilean relativistic symmetries could no longer be upheld.
From a modern perspective we should see the pre-Einsteinian attempts to address that
crisis (such as the ones of Lorentz) as attempts to ``break Galilean invariance",
 {\it i.e.} preserve the validity of Galilean transformations
as laws of transformation among inertial observers, but renouncing to the possibility that those
transformations be a symmetry of the laws of physics. The ``ether" would be a preferred frame
for the description of the laws of physics, and the laws of physics that hold in other frames
would be obtained from the ones of the preferred frame via Galilean transformations.
Those attempts failed.\\
What succeeded is completely complementary. Experimental evidence, and the analyses
of Einstein (and Poincar\'e) led us to a ``deformation of Galilean invariance":
in Special Relativity the laws of transformation
among observers still are a symmetry of the laws of physics (Special Relativity is no less
relativistic then Galilean Relativity), but the special-relativistic transformation laws
are a $c$-deformation of the Galilean laws of transformation with the special property
of achieving the observer-independence of the speed scale $c$.\\
This famous $c$-deformation in particular replaces the Galilean on-shell relation
$E= \mathrm{constant} + {\bf p}^2/(2m)$ with the special-relativistic version
$E= \sqrt{c^2{\bf p}^2+c^4 m^2}$
and the Galilean
composition of velocities ${\bf u} \oplus {\bf v} = {\bf u} + {\bf v}$
with the much more complex special-relativistic law of composition of velocities.\\
This interplay between $c$-deformation of Galilean transformations and the associated
deformations of the law of composition of velocities, is analogous to the
interplay between the DSR-type $\ell$-deformation of Lorentz transformations
and the associated deformations of the law of composition of momenta.

\section{A known example of DSR setup with universality}\label{kappauno}
I shall now give more tangibility to the brief review of DSR concepts contained
in the previous section, by discussing a known DSR setup and highlighting
the connection between deformation of the on-shell relation and deformation
of the laws of momentum conservation.\\
The specific DSR setup reviewed in this section will also provide the starting
point for the generalization introduced in this manuscript. The one that I review
in this section, following mainly the results of Ref.~\cite{goldenrule},
still is a standard DSR setup with ``universal effects", {\it i.e.} the deformation
of Lorentz symmetry affects all particles in exactly the same way.
Then in the next sections I will take the DSR setup of this section as starting
point for adding the possibility of ``nonuniversal effects", {\it i.e.}
cases where the deformation
of Lorentz symmetry affects different types of particles in different ways.

The DSR setup on which I focus in this section
was analyzed from the perspective here relevant
in the recent Ref.~\cite{goldenrule}, and more preliminarily in previous DSR studies
(some comments on it were already in Ref.~\cite{dsr1Edsr2}).
It is centered on a choice of on-shell relation and law of composition of momenta
which became recently of interest\cite{flagiuKAPPAPRL,anatomy}
also in the study of the new proposal
of ``relative-locality momentum spaces"~\cite{prl,grf2nd},
and provides a set of rules for kinematics which can be naturally described
from the viewpoint of the $\kappa$-Poincar\'e Hopf algebra~\cite{majidruegg,kpoinap}.

The on-shell (``dispersion") relation is (for the 1+1-dimensional case)
\begin{equation}
m^2 = p_0^2 - p_1^2 + \ell p_0 p_1^2
\label{metrictorsy}
\end{equation}
and the law of composition of momenta is
\begin{equation}
(p \oplus_\ell p^\prime)_1 = p_1 + p^\prime_1 + \ell p_0 p^\prime_1  ~,~~~
(p \oplus_\ell p^\prime)_0 = p_0 + p^\prime_0~.
\label{connectiontorsy}
\end{equation}

%I note down the antipode for the MR composition law:
%\begin{equation}
%(\ominus p)_j = - p_j + \ell p_0 p_j  ~,~~~
%(\ominus p)_0 = - p_0~,
%\label{antipodetorsy}
%\end{equation}
%which indeed, as verified by direct application of (\ref{connectiontorsy}),
%is such that $p \oplus (\ominus p) = 0$.\\
%And I also observe that from (\ref{connectiontorsy}) it follows that
%\begin{equation}
%[(k \oplus p) \oplus q]_1 = k_1 + p_1 + q_1 +  \ell k_0 p_1
%+  \ell k_0 q_1 +  \ell p_0 q_1  ~,~~~
%[(k \oplus p) \oplus q]_0 = k_0 + p_0 + q_0
%\label{tribodytorsy}
%\end{equation}

The on-shell relation (\ref{metrictorsy}) is invariant
under the following action of a boost on the momentum
of a particle:
\begin{equation}
[N, p_0] =  p_1 ~,~~~
[N, p_1] =  p_0 + \ell p_0^2 + \frac{\ell}{2} p_1^2~,
\label{booststorsy}
\end{equation}
which indeed ensures
\begin{equation}
[N, p_0^2 - p_1^2 +  \ell p_0 p_1^2] =
2 p_0 p_1 - 2 p_1 (p_0 + \ell p_0^2 + \frac{\ell}{2} p_1^2)
+\ell p_1^3 +2 \ell p_0^2 p_1 = 0
\label{invariantshelltorsy}
\end{equation}
In light of the points highlighted in the brief review of DSR concepts
offered in the previous section it should be clear that this invariance
of the on-shell relation would not in itself establish the DSR-compatibility
of this setup: we must also insist that the laws of conservation of momentum,
written using the composition law (\ref{connectiontorsy}), are covariant
under the action of the boosts that leave the on-shell relation invariant.\\
And this requirement of covariance of conservation laws is rather challenging
when the law of composition of momenta is non-commutative,
as in the case of (\ref{connectiontorsy}). I am not advocating that the law of composition
of momenta {\underline{should}} be non-commutative; on the contrary one may well
prefer~\cite{goldenrule} commutative laws of composition of momenta in a DSR setup.
But part of the strength of the results I am here reporting resides in the
fact that I am able to generalize to ``nonuniversal DSR deformations" not
merely a particularly simple DSR setup previously known with universality,
but actually a rather virulent DSR setup, which in particular relies on
a noncommutative law of composition of momenta. This should reassure my readers of the
fact that the new structures I introduce in this manuscript should allow to produce
non-universal versions of a rather large variety of DSR setups.

However, as stressed and addressed in Ref.~\cite{goldenrule},
the non-commutativity of the composition law (\ref{connectiontorsy})
 evidently poses a challenge for formulating the action of boosts on
momenta obtained composing two or more single-particle momenta.
In Special Relativity (and in DSR setups with commutative law of composition
of momenta~\cite{goldenrule})
 it is possible to
simply impose that the boost of a two-particle
event $e_{p \oplus_\ell p^\prime}$ be governed by
$$[N_{[p]}+N_{[p^\prime]},p \oplus_\ell p^\prime] $$
where  I decomposed
the action of boosts on the composed momentum into two pieces, each given in terms
of a boost acting exclusively on a certain momentum in the event.\\
This means that in Special Relativity one has
 a ``total boost generator" obtained by combining trivially
the boost generators acting on each individual particle.
But with a noncommutative law of composition of momenta this simplicity is lost:
the lack of symmetry under exchange of particles
precludes, as
 one can easily verify,
  the possibility of adopting
a ``total boost generator" given by a trivial sum of single-particle
boost generators.
There is in particular no choice~\cite{goldenrule} of $N_{[p]}$ capable of ensuring
that $[N_{[p]}+N_{[p^\prime]}+N_{[p^{\prime\prime}]},(p \oplus_\ell p^{\prime}
 \oplus_\ell p^{\prime\prime})_\mu] $ vanishes
whenever $(p \oplus_\ell p^{\prime} \oplus_\ell p^{\prime\prime})_\mu =0$.

What does work, as shown in Ref.~\cite{goldenrule}, is adopting a corresponding
deformation of the ``total-boost law"
\begin{eqnarray}
N_{[p \oplus_\ell p^{\prime}]} =N_{[p]}+N_{[p^{\prime}]} + \ell p_0 N_{[p^{\prime}]}
\label{boosttwo}
\end{eqnarray}
and accordingly
\begin{eqnarray}
N_{[p \oplus_\ell p^{\prime} \oplus_\ell p^{\prime\prime}]}
=N_{[p]}+N_{[p^{\prime}]}+N_{[p^{\prime\prime}]} + \ell p_0 N_{[p^{\prime}]}
+ \ell p_0 N_{[p^{\prime\prime}]}
+\ell p^{\prime}_0 N_{[p^{\prime\prime}]}~.
\label{boostthree}
\end{eqnarray}

In Ref.~\cite{goldenrule} I verified in some detail that this prescription
produces a fully consistent relativistic framework, with the needed
compatibility between on-shell relation (\ref{metrictorsy})
and law of composition of momenta (\ref{connectiontorsy}):
the on-shell relation is invariant and the laws of conservation of momentum
obtained from the composition law are covariant.\\
Let me here just review briefly the specific result of Ref.~\cite{goldenrule}
concerning the covariance of the conservation law for a ``trivalent process"
with\footnote{{\bf Throughout this manuscript
I write conservation laws at a process with conventions such that
all momenta intervening in the process are incoming into the process, so that
indeed a trivalent process would be characterized by a conservation
law of the type $p \oplus_\ell p^{\prime} \oplus_\ell p^{\prime\prime} = 0$.
The case of one (or two) of the momenta that is outgoing from the process,
say the momentum $p$,
is recovered by simply substituting for $p$ the ``antipode" of the momentum of
that outgoing particle,
with the antipode $\ominus p$ defined so that $(\ominus p) \oplus p =0$.
[For the composition law (\ref{connectiontorsy})
the antipode is such that $(\ominus_\ell p)_j = - p_j + \ell p_0 p_j$
and $(\ominus_\ell p)_0 = - p_0$.]}}
$p \oplus_\ell p^{\prime} \oplus_\ell p^{\prime\prime} = 0$.\\
Checking that $N_{[p]}+N_{[p^{\prime}]}+N_{[p^{\prime\prime}]} + \ell p_0 N_{[p^{\prime}]}
+ \ell p_0 N_{[p^{\prime\prime}]} +\ell p^{\prime}_0 N_{[p^{\prime\prime}]}$ does indeed ensure
the relativistic covariance of
the conservation law $p \oplus_\ell p^{\prime} \oplus_\ell p^{\prime\prime} = 0$ is best
done considering separately the $0$ (``time") component
and the spatial $1$ (``spatial") component.
For the $0$ component one easily finds~\cite{goldenrule}
\begin{eqnarray}
&& \!\!\!\!\!\!\!\!\!\!\!\!\!\!\!\!\!\!\!\!\!\!\!\!\!\!\!\!\!\!\!\!\!
[N_{[p]}+N_{[p^{\prime}]}+N_{[p^{\prime\prime}]} + \ell p_0 N_{[p^{\prime}]}
+ \ell p_0 N_{[p^{\prime\prime}]}
+\ell p^{\prime}_0 N_{[p^{\prime\prime}]}, (p \oplus_\ell p^{\prime} \oplus_\ell p^{\prime\prime})_0]
 =
 \nonumber\\
 && =  [N_{[p]}+N_{[p^{\prime}]}+N_{[p^{\prime\prime}]} + \ell p_0 N_{[p^{\prime}]}
+ \ell p_0 N_{[p^{\prime\prime}]}
+\ell p^{\prime}_0 N_{[p^{\prime\prime}]}, p_0 + p^{\prime}_0 + p^{\prime\prime}_0 ]=\nonumber\\
 \!\!\!\!\!\!\! \!\!\!\!\!\!\! \!\!\!\!\!\!\!
 & & = p_1+ p^{\prime}_1 +p^{\prime\prime}_1+
 \ell p_0 p^{\prime}_1 + \ell p_0 p^{\prime\prime}_1 + \ell p^{\prime}_0 p^{\prime\prime}_1 =
 (p \oplus_\ell p^{\prime} \oplus_\ell p^{\prime\prime})_1 = 0~,
\label{covatritimetorsy}
\end{eqnarray}
where on the right-hand side I of course used the conservation law itself.\\
Similarly for the $1$ component one easily finds
that~\cite{goldenrule}
\begin{eqnarray}
&& \!\!\!\!\!\!\!\!\!
[N_{[p]}+N_{[p^{\prime}]}+N_{[p^{\prime\prime}]} + \ell p_0 N_{[p^{\prime}]}
+ \ell p_0 N_{[p^{\prime\prime}]}
+\ell p^{\prime}_0 N_{[p^{\prime\prime}]},  (p \oplus_\ell p^{\prime} \oplus_\ell p^{\prime\prime})_1]
 = \nonumber\\
 && =  [N_{[p]}+N_{[p^{\prime}]}+N_{[p^{\prime\prime}]} + \ell p_0 N_{[p^{\prime}]}
+ \ell p_0 N_{[p^{\prime\prime}]}
+\ell p^{\prime}_0 N_{[p^{\prime\prime}]},
p_1+ p^{\prime}_1 +p^{\prime\prime}_1+
 \ell p_0 p^{\prime}_1 + \ell p_0 p^{\prime\prime}_1 + \ell p^{\prime}_0 p^{\prime\prime}_1]=\nonumber\\
 & &
=  p_0+\ell p_0^2 + \frac{\ell}{2} p_1^2
+ p^{\prime}_0+\ell p^{\prime \, 2}_0  + \frac{\ell}{2} p^{\prime \, 2}_1  +
 p^{\prime\prime}_0+\ell p^{\prime\prime \, 2}_0 + \frac{\ell}{2} p^{\prime\prime \, 2}_1
+ 2 \ell p_0 p^{\prime}_0 + 2 \ell p_0 p^{\prime\prime}_0 + 2 \ell p^{\prime}_0 p^{\prime\prime}_0
+  \ell p_1 p^{\prime}_1 +  \ell p_1 p^{\prime\prime}_1
+\ell p^{\prime}_1 p^{\prime\prime}_1 = \nonumber\\
&&= p_0+p^{\prime}_0 +p^{\prime\prime}_0
+\ell  (p_0 + p^{\prime}_0 + p^{\prime\prime}_0)^2
+ \frac{\ell}{2} (p_1 + p^{\prime}_1 +p^{\prime\prime}_1)^2 = 0
~~~~~~~~~~~~~~~~~~~~~~~~~~~~~~~~~~~~~~~~~~~~~~~~~~~~~
\label{covatrispacetorsy}
\end{eqnarray}
where again on the right-hand side I
used the conservation law $p \oplus_\ell p^{\prime} \oplus_\ell p^{\prime\prime} = 0$ itself,
and I took again into account that
I am working at leading order in $\ell$.

The results (\ref{covatritimetorsy}) and (\ref{covatrispacetorsy})
establish that indeed the boosts
(\ref{booststorsy}),(\ref{boosttwo})(\ref{boostthree}),
besides admitting the on-shell relation (\ref{metrictorsy})
as invariant,
also admit $p \oplus_\ell p^{\prime} \oplus_\ell p^{\prime\prime} = 0$
as a covariant law.

\section{A first example of DSR setup without universality}\label{hopfliesec}
The previous sections only provided the preliminaries for the main analysis
that I report in this manuscript, which is contained in this and the next section.
It was a rather bulky effort on preliminaries, but I felt this might
be beneficial since the
results I am reporting are to a large extent suprising/unexpected and it might
be helpful for my readers to be equipped with a nearly self-contained summary
of the previous results which provide the starting point for the analysis I am here reporting.
As announced, I am going to show that
there are logically consistent DSR-relativistic
theories in which different types of particles
are governed by different laws of kinematics.\\
In this section I start by establishing that the type of particle discussed in the
previous section, governed
by a specific DSR $\ell$-deformation of Lorentz symmetry,
can coexist with a second type of particle, governed by ordinary Special Relativity.\\
The key point will be to show that there are laws of conservation of momentum
allowing momentum to be exchanged between the two types of particles\footnote{If the two
types of particles could not interact then they  would actually not ``coexist": the ``Universe"
of $\ell$-deformed particles would remain decoupled from (undetectable and irrelevant for)
the ``Universe"
of $\ell$-deformed particles, and {\it vice versa}.}
in a fully relativistic manner.

I shall consistently denote with $p$ (or $p^{\prime}$ or $p^{\prime\prime}$ ...)
the momenta of the type of
particles affected by the
DSR $\ell$-deformation of Lorentz symmetry discussed in the previous section,
so that in particular
\begin{equation}
m^2 = p_0^2 - p_j^2 + \ell p_0 p_j^2~.
\label{metricHLp}
\end{equation}
And I shall consistently denote with $k$ (or $k^{\prime}$ or $k^{\prime\prime}$ ...)
the momenta of particles of the second type, the type governed
by undeformed Special Relativity,
so that in particular
\begin{equation}
\mu^2 = k_0^2 - k_j^2
\label{metricHLk}
\end{equation}

For the first type (``$p$-type") of particles I shall insist again on
\begin{equation}
[N_{[p]}, p_0] =  p_1 ~,~~~
[N_{[p]}, p_1] =  p_0 + \ell p_0^2 + \frac{\ell}{2} p_1^2~,
\label{boostsHL}
\end{equation}
while naturally, for the second type (``$k$-type") of particles, I take
\begin{equation}
[N_{[k]}, k_0] =  k_1 ~,~~~
[N_{[k]}, k_1] =  k_0 ~.
\label{boostSRhl}
\end{equation}

And naturally the composition law for the special-relativistic momenta (``$k$-type")
is standard,
\begin{equation}
(k \oplus_{\star} k')_j = k_j + k'_j   ~,~~~
(k \oplus_{\star} k')_0 = k_0 + k'_0~,
\label{connectionHLk}
\end{equation}
whereas
the composition law for the $\ell$-deformed (``$p$-type") particles
was already introduced in the previous section:
\begin{equation}
(p \oplus_{\ell} p^\prime)_j = p_j + p^\prime_j + \ell p_0 p^\prime_j  ~,~~~
(p \oplus_{\ell} p^\prime)_0 = p_0 + p^\prime_0~.
\label{connectionHLp}
\end{equation}

It is easy to see that the main challenge resides in finding a consistent
way to compose momenta of different types, {\it i.e.}
finding some ``mixing composition law", of the type $p \oplus_{\ell \star} k$, while insisting
that conservation laws written in terms of such a composition law would be covariant
under a consistent prescription for the action of boosts.
Equipped with no theorem, but rather the findings of a lengthy ``trial and error exercise"
%(actually nearly a decade's worth of a few such trials and very many errors)
I can simply exhibit an example of such a ``mixing composition law" which does work,
and it is remarkably simple:
\begin{equation}
(p \oplus_{\ell \star} k)_j = p_j + k_j + \frac{\ell}{2} p_0 k_j  ~,~~~
(p \oplus_{\ell,\star} k)_0 = p_0 + k_0~.
\label{connectionHLpk}
\end{equation}

In order to convince my readers that this does work let me start slowly,
focusing at first on ``bi-valent processes"  in this theory
with two types of particles.
Some of these bi-valent processes, the ones without ``mixing",
will pose no challenge:
processes with
conservation law  $k \oplus_{\star} k^\prime =  0$
will have covariance ensured by standard total-boost actions
\begin{equation}
N_{[k \oplus_{\star} k']} =N_{[k]}+N_{[k']}
\label{boostkk}
\end{equation}
while processes with
conservation law  $p \oplus_{\ell} p^\prime =  0$
will have covariance ensured by the total-boost actions discussed
in the previous section
\begin{equation}
N_{[p \oplus_{\ell} p^\prime]} =N_{[p]}+N_{[p^\prime]} + \ell p_0 N_{[p^\prime]}~.
\label{boostpp}
\end{equation}
The only class of bi-valent processes which involves ``mixing"
is the one of processes
with conservation law of the type $p \oplus_{\ell \star} k=0$.
In light of the discussion offered in the previous section it should
not be too surprising that a relativistically consistent description
of such ``mixing bi-valent processes" is obtained in terms
of a total-boost action which itself involves a sort of ``mixing":
\begin{equation}
N_{[p \oplus_{\ell \star} k]} =N_{[p]}+N_{[k]} + \frac{\ell}{2} p_0 N_{[k]}~.
\label{boostpk}
\end{equation}
Let us verify that indeed this total-boost action
ensure the covariance of the
conservation law $p \oplus_{\ell \star} k=0$.
I start again from the $0$ component of $p \oplus_{\ell \star} k=0$
for which I find
\begin{equation}
[N_{[p]}+N_{[k]} + \frac{\ell }{2} p_0 N_{[k]}, k_0 + p_0]
= p_1+k_1 + \frac{\ell}{2} p_0 k_1 =
(p \oplus_{\ell \star} k)_1= 0
\label{covatwozeroHLpk}
\end{equation}
where on the right-hand side I of course
used again the conservation law itself.\\
Similarly for the $1$ component of $p \oplus_{\ell \star} k=0$ I find
\begin{eqnarray}
&& [N_{[p]}+N_{[k]} + \frac{\ell}{2} p_0 N_{[k]},
p_1+k_1 + \frac{\ell }{2} p_0 k_1 ]
 = p_0+\ell p_0^2 + \frac{\ell}{2} p_1^2+ k_0
  + \frac{\ell}{2} p_1 k_1 + \ell p_0 k_0 =
  \label{covatwounoHLpk}\\
&& ~~~~~  =  p_0+k_0
+ \ell p_0 (p_0 + k_0)
 + \frac{\ell}{2} p_1 (p_1 + k_1) = 0
\nonumber
\end{eqnarray}
where again on the right-hand side I
used the conservation law $p \oplus_{\ell \star} k=0$
and took again into account that
I am working at leading order in $\ell$.
[Working at leading order in $\ell$ one
finds, {\it e.g.}, $\ell p_1 [(p \oplus_{\ell \star} k)_1] \simeq
\ell p_1 [p_1  + k_1] $.]

So I did find that my description of boosts and of the composition
laws $\oplus_{\ell \star} , \oplus_{\ell} , \oplus_{\star}$
ensures the covariance of conservation laws for  all ``bi-valent processes".

I can now move on to the case of tri-valent processes. The covariance of
the kinematics of processes
with $k \oplus_{\star} k' \oplus_{\star} k^{\prime\prime}=0$
and $p \oplus_{\ell} p^\prime \oplus_{\ell} p^{\prime\prime}=0$
is assured respectively by Special Relativity
and by the results reviewed in the previous section.
Of course the only potentially troublesome tri-valent processes
are the ones whose conservation laws involve ``mixing",
which leads me to focus on the cases $p \oplus_{\ell} p^\prime \oplus_{\ell \star} k=0$
and $p \oplus_{\ell  \star} k \oplus_{\star} k'=0$.\\
In order to show that these ``tri-valent mixing conservation laws"
are covariant under the action of the boosts
$$N_{[p]}+N_{[p^\prime]}+N_{[k]} + \ell  p_0 N_{[p^\prime]}
+ \frac{\ell}{2} (p_0 + p_0') N_{[k]}$$
let me start from the $0$ component of $p \oplus_{\ell} p^\prime \oplus_{\ell \star} k=0$, for
which I easily find the expected result
\begin{eqnarray}
&& [N_{[p]}+N_{[p^\prime]}+N_{[k]} + \ell  p_0 N_{[p^\prime]}
+ \frac{\ell}{2} (p_0 + p_0') N_{[k]},
p_0 + p^\prime_0 + k_0]
= p_1+p^\prime_1+k_1 + \ell  p_0 p^\prime_1 + \frac{\ell}{2} (p_0 + p^\prime_0) k_1 =\nonumber\\
&& ~~~~~~~~~~~~~~~~~  =(p \oplus_{\ell} p^\prime \oplus_{\ell \star} k)_1= 0
\label{covathreezeroHLppk}
\end{eqnarray}
Also successful is the verification
for the $1$ component of  $p \oplus_{\ell} p^\prime \oplus_{\ell \star} k=0$,
which progresses as follows:
\begin{eqnarray}
&& [N_{[p]}+N_{[p^\prime]}+N_{[k]} + \ell  p_0 N_{[p^\prime]}
+ \frac{\ell}{2} (p_0 + p_0') N_{[k]},
p_1+p^\prime_1+k_1 + \ell  p_0 p^\prime_1 + \frac{\ell}{2} (p_0 + p^\prime_0) k_1  ] =\nonumber\\
&& ~~~~~~~~ = p_0+\ell p_0^2 + \frac{\ell}{2} p_1^2
+ p^\prime_0+\ell p^{\prime \, 2}_0 + \frac{\ell}{2} p^{\prime \, 2}_1 + k_0 +\nonumber\\
&& ~~~~~~~~ ~~~~  + \ell p_1 p^\prime_1 + 2 \ell p_0 p^\prime_0
+ \frac{\ell}{2} (p_1 + p^\prime_1) k_1 + \ell (p_0 + p^\prime_0) k_0 =
 \nonumber\\
&& ~~~~~  =  p_0+ p^{\prime}_0+k_0
+ \ell (p_0 + p^{\prime}_0) (p_0 + p^{\prime}_0 + k_0)
 + \frac{\ell}{2} (p_1 + p^{\prime}_1) (p_1 + p^{\prime}_1 + k_1)= 0
 \label{covathreeunoHLppk}
\end{eqnarray}
Having done this it is not hard to adapt the results I obtained
for $p \oplus_{\ell} p^\prime \oplus_{\ell \star} k=0$ to the
only slightly different case of $p \oplus_{\ell \star} k \oplus_{\star} k'=0$.
In order to establish also the covariance of $p \oplus_{\ell  \star} k \oplus_{\star} k'=0$
let me start again with its $0$ component, for which I easily find a satisfactory
 result:
\begin{eqnarray}
&& [N_{[p]}+N_{[k]} +N_{[k^\prime]}+ \frac{\ell}{2} p_0  ( N_{[k]} +  N_{[k^\prime]}),
p_0  + k_0 + k^\prime_0]
= p_1+k_1 +k^\prime_1
+ \frac{\ell}{2} p_0  (k_1 + k^\prime_1)   =\nonumber\\
&& ~~~~~~~~~~~~~~~~~  =(p \oplus_{\ell \star} k \oplus_{\star} k')_1= 0
\label{covathreezeroHLpkkB}
\end{eqnarray}
And equally satisfactory  is the corresponding
 verification
for the $1$-component of $p \oplus_{\ell , \lambda} k \oplus_{\lambda} k'=0$
which progresses as follows:
\begin{eqnarray}
&& [N_{[p]}+N_{[k]} +N_{[k^\prime]}+ \frac{\ell}{2} p_0  ( N_{[k]} +  N_{[k^\prime]}),
p_1+k_1 + k^\prime_1
+ \frac{\ell}{2} p_0  (k_1 + k^\prime_1) ] =\nonumber\\
&& ~~~~~~~~ = p_0+\ell p_0^2 + \frac{\ell}{2} p_1^2
+ k_0
+ k^\prime_0
+ \frac{\ell}{2} p_1  (k_1 + k^\prime_1) + \ell p_0 ( k_0+ k^\prime_0)=
 \nonumber\\
&& ~~~~~~~~ = p_0
+ k_0
+ k^\prime_0
+ \frac{\ell}{2} p_1  (p_1 + k_1 + k^\prime_1) + \ell p_0 (p_0 + k_0+ k^\prime_0)= 0
 \label{covathreeunoHLpkkB}
\end{eqnarray}

So I did show that there is at least one (and surely many more) example
of ``mixing interaction" which satisfies the demands of the relativity of inertial frames,
producing rules of relativistic kinematics with a consistent description of interactions
among particles with different relativistic properties:
the results (\ref{covathreezeroHLppk}), (\ref{covathreeunoHLppk}), (\ref{covathreezeroHLpkkB}), (\ref{covathreeunoHLpkkB})
confirm that the boosts I introduced in
(\ref{boostsHL}), (\ref{boostSRhl}),
(\ref{boostkk}), (\ref{boostpp}), (\ref{boostpk}),
besides admitting the on-shell relations
(\ref{metricHLp}),(\ref{metricHLk})
as invariants,
also admit the conservation laws
$p \oplus_\ell p^{\prime} \oplus_\ell p^{\prime\prime} = 0$
and $p \oplus_{\ell  \star} k \oplus_{\star} k'=0$
as covariant laws.\\
This establishes that it is possible to have meaningful (interacting)
theories that are fully relativistic in spite of allowing for the
coexistence of a type of particle whose relativistic properties
are governed by ordinary Special Relativity
and of a type of particle whose relativistic properties
are governed
by the type of DSR $\ell$-deformation of Lorentz symmetry reviewed in the previous section.

\section{From Hopf-Lie to Hopf-Hopf}\label{hopfhopfsec}
The new results reported in the previous section
establish a possibility for coexistence of DSR-relativistic
and ordinarily special-relativistic particles. As I shall stress in later parts
of this manuscript, this should also invite
(in addition to possibly other schemes of mathematical implementation)
 the development of mathematical structures
suitable for ``mixing" the structure of a non-trivial Hopf algebra and the
structure of a (Hopf algebra with primitive coproducts, {\it i.e.} a)
Lie algebra. I am now going to report results that further generalize the realm
of possibilities: the case of two types of particles, with both types governed
by the specific DSR deformation of Lorentz symmetry reviewed in Sec.~\ref{kappauno},
but one type of particle has DSR deformation scale $\ell$ while the other type
of particle has DSR deformation scale $\lambda$ (so that one could describe the framework
as mixing the ``Hopf-algebra properties" of one type of particles with the somewhat
different ``Hopf-algebra properties" of another type of particles). This will allow me,
in later parts
of this manuscript, to speculate about the possibility that $\lambda$ for a ``composite particle"
might be obtained from $\ell$ via some rescaling law possibly based on the number and nature
of the constituents (an illustrative example of which would be $\lambda = \ell/N$, with
the $N$ the number of constituents).

I shall consistently denote with $p$ (or $p^{\prime}$ or $p^{\prime\prime}$ ...)
the momenta of the type of
particles affected by the
DSR $\ell$-deformation of Lorentz symmetry discussed Sec.~\ref{kappauno},
so that in particular
\begin{equation}
m^2 = p_0^2 - p_j^2 + \ell p_0 p_j^2
\label{metricHHp}
\end{equation}
And I shall consistently denote with $k$ (or $k^{\prime}$ or $k^{\prime\prime}$ ...)
the momenta of particles of the second type, the type
with DSR $\lambda$-deformation of Lorentz symmetry,
so that in particular
\begin{equation}
\mu^2 = k_0^2 - k_j^2 + \lambda k_0 k_j^2
\label{metricHHk}
\end{equation}

Of course, for boosts acting on a single-particle momentum I will rely again on
\begin{equation}
[N_{[p]}, p_0] =  p_1 ~,~~~
[N_{[p]}, p_1] =  p_0 + \ell p_0^2 + \frac{\ell}{2} p_1^2~,
\label{boostsHHp}
\end{equation}
and accordingly
\begin{equation}
[N_{[k]}, k_0] =  k_1 ~,~~~
[N_{[k]}, k_1] =  k_0 + \lambda k_0^2 + \frac{\lambda}{2} k_1^2~.
\label{boostHHk}
\end{equation}

And also for the ``laws of composition without mixing", $p \oplus_{\ell} p^\prime$
and $k \oplus_{\lambda} k^\prime$,
everything is clear from the onset of the analysis:
\begin{equation}
(p \oplus_{\ell} p^\prime)_j = p_j + p^\prime_j + \ell p_0 p^\prime_j  ~,~~~
(p \oplus_{\ell} p^\prime)_0 = p_0 + p^\prime_0~.
\label{connectionHHp}
\end{equation}
\begin{equation}
(k \oplus_{\lambda} k')_j = k_j + k'_j + \lambda k_0 k'_j  ~,~~~
(k \oplus_{\lambda} k')_0 = k_0 + k'_0~.
\label{connectionHHk}
\end{equation}
In particular, this ensures again that processes involving
only ``laws of composition without mixing" will be described in consistent relativistic
manner by enforcing the ``laws of action of boosts without mixing" which I already used above:
$$N_{[p \oplus_{\ell} p^\prime]} =N_{[p]}+N_{[p^\prime]} + \ell p_0 N_{[p^\prime]}$$
$$N_{[k \oplus_{\lambda} k']} =N_{[k]}+N_{[k']} + \lambda k_0 N_{[k']}$$

So once again the main challenge resides in finding a consistent
way to compose momenta of different types of particles, {\it i.e.}
finding some ``mixing composition law" $p \oplus_{\ell , \lambda} k$, while insisting
that conservation laws written in terms of such a composition law would be covariant
under a consistent prescription for the action of boosts.
Relying on the findings of a lengthy ``trial and error exercise"
I can simply exhibit an example of such a ``mixing composition law" which does work,
and it is once again remarkably simple:
\begin{equation}
(p \oplus_{\ell,\lambda} k)_j = p_j + k_j + \frac{\ell + \lambda}{2} p_0 k_j  ~,~~~
(p \oplus_{\ell,\lambda} k)_0 = p_0 + k_0~.
\label{connectionHHpk}
\end{equation}
And I shall also show that the needed counterpart taking the shape of a ``mixing composition
of boosts" that leads to a consistently relativistic description is given by
$$N_{[p \oplus_{\ell,\lambda} k]} =N_{[p]}+N_{[k]} + \frac{\ell + \lambda}{2} p_0 N_{[k]}$$

With these characterizations I have completed the specifications of kinematics needed for
the results being reported in this section. I shall now proceed to deriving these results.
Readers are invited to notice that all specifications and results given in this section
evidently reduce to the ones of the previous section in the limit $\lambda \rightarrow 0$.

Let us verify that indeed these boost actions
ensure compatibility with the
conservation laws obtained from the composition laws $\oplus_{\ell}$,
$\oplus_{\lambda}$, and
$\oplus_{\ell,\lambda}$. Actually for processes involving either exclusively
$\oplus_{\ell}$, or exclusively
$\oplus_{\lambda}$ the compatibility with my deformed boosts was already verified in
Sec.~\ref{kappauno}. So I can focus on cases which involve at least one $\oplus_{\ell,\lambda}$.
Looking first at ``bi-valent processes" the only case of interest
then evidently is $p \oplus_{\ell,\lambda} k=0$.
For the $0$-th component of $p \oplus_{\ell,\lambda} k=0$ one finds
\begin{equation}
[N_{[p]}+N_{[k]} + \frac{\ell + \lambda}{2} p_0 N_{[k]}, k_0 + p_0]
= p_1+k_1 + \frac{\ell + \lambda}{2} p_0 k_1 =
(p \oplus_{\ell,\lambda} k)_1= 0
\label{covatwozeroHHpk}
\end{equation}
where on the right-hand side I of course
used again the conservation law itself.\\
Similarly for the $1$-component of $p \oplus_{\ell,\lambda} k=0$ one finds
\begin{eqnarray}
&& [N_{[p]}+N_{[k]} + \frac{\ell + \lambda}{2} p_0 N_{[k]},
p_1+k_1 + \frac{\ell + \lambda}{2} p_0 k_1 ]
 = p_0+\ell p_0^2 + \frac{\ell}{2} p_1^2+ k_0+\lambda k_0^2  + \frac{\lambda}{2} k_1^2
  + \frac{\ell + \lambda}{2} p_1 k_1 + (\ell + \lambda) p_0 k_0 =
 \nonumber\\
&& ~~~~~ = p_0 + k_0
 + \frac{\ell - \lambda}{2} (p_0^2 - k_0^2) + \frac{\ell - \lambda}{4} (p_1^2 - k_1^2)+
 \frac{\ell + \lambda}{2} (p_0^2 + k_0^2 +2 k_0 p_0)
 + \frac{\ell + \lambda}{4} (p_1^2 + k_1^2 +2 k_1 p_1) =  \label{covatwounoHHpk}\\
&& ~~~~~  =  p_0+k_0
+ \frac{\ell - \lambda}{2} (p_0 - k_0) (p_0 + k_0)
 + \frac{\ell - \lambda}{4} (p_1 - k_1) (p_1 + k_1)
 +  \frac{\ell + \lambda}{2} (p_0 + k_0)^2
 + \frac{\ell + \lambda}{4} (p_1 + k_1)^2 = 0
\nonumber
\end{eqnarray}
where again on the right-hand side I
used the conservation law $p \oplus_{\ell,\lambda} k=0$
and took again into account that
I am working at leading order in $\ell,\lambda$.
[Working at leading order in $\ell,\lambda$ one
finds $(\ell + \lambda) [(p \oplus_{\ell,\lambda} k)_1]^2 \simeq (\ell + \lambda) (p_1 + k_1)^2$
and $(\ell - \lambda) (p_1 - k_1) (p \oplus_{\ell,\lambda} k)_1
 \simeq (\ell - \lambda) (p_1 - k_1) (p_1 + k_1)$.]

So I did find that my description of boosts (and of the composition
laws $\oplus_{\ell,\lambda} , \oplus_{\ell} , \oplus_{\lambda}$)
ensures the convariance of conservation laws for  ``bi-valent processes".

I can now move on to the case of tri-valent processes. The covariance of
the kinematics of processes
with $p \oplus_{\ell} p^\prime \oplus_{\ell} p^{\prime\prime}=0$
 and $k \oplus_{\lambda} k' \oplus_{\lambda} k^{\prime\prime}=0$
is already assured by the results here reviewed in Sec.~\ref{kappauno}.
I shall now verify that the formulation of boost transformations here introduced
affords me the covariance also of the cases with $p \oplus_{\ell} p^\prime \oplus_{\ell , \lambda} k=0$
and $p \oplus_{\ell , \lambda} k \oplus_{\lambda} k'=0$.\\
Let me start from the $0$-th component of $p \oplus_{\ell} p^\prime \oplus_{\ell,\lambda} k=0$ for
which I easily find the expected result
\begin{eqnarray}
&& [N_{[p]}+N_{[p^\prime]}+N_{[k]} + \ell  p_0 N_{[p^\prime]}
+ \frac{\ell + \lambda}{2} (p_0 + p_0') N_{[k]},
p_0 + p^\prime_0 + k_0]
= p_1+p^\prime_1+k_1 + \ell  p_0 p^\prime_1 + \frac{\ell + \lambda}{2} (p_0 + p^\prime_0) k_1 =\nonumber\\
&& ~~~~~~~~~~~~~~~~~  =(p \oplus_{\ell} p^\prime \oplus_{\ell,\lambda} k)_1= 0
\label{covathreezeroHHppk}
\end{eqnarray}
Also successful, but slightly more tedious, is the verification
of the covariance for the $1$-component of  $p \oplus_{\ell} p^\prime \oplus_{\ell,\lambda} k=0$,
which progresses as follows:
\begin{eqnarray}
&& [N_{[p]}+N_{[p^\prime]}+N_{[k]} + \ell  p_0 N_{[p^\prime]}
+ \frac{\ell + \lambda}{2} (p_0 + p_0') N_{[k]},
p_1+p^\prime_1+k_1 + \ell  p_0 p^\prime_1 + \frac{\ell + \lambda}{2} (p_0 + p^\prime_0) k_1  ] =\nonumber\\
&& ~~~~~~~~ = p_0+\ell p_0^2 + \frac{\ell}{2} p_1^2
+ p^\prime_0+\ell p^{\prime \, 2}_0 + \frac{\ell}{2} p^{\prime \, 2}_1 + k_0+\lambda k_0^2  + \frac{\lambda}{2} k_1^2 +\nonumber\\
&& ~~~~~~~~ ~~~~  + \ell p_1 p^\prime_1 + 2 \ell p_0 p^\prime_0
+ \frac{\ell + \lambda}{2} (p_1 + p^\prime_1) k_1 + (\ell + \lambda) (p_0 + p^\prime_0) k_0 =
 \nonumber\\
&& ~~~~~ = p_0 +p^\prime_0 + k_0
 + \frac{\ell - \lambda}{2} (p_0^2 + p^{\prime \, 2}_0 - k_0^2)
 + \frac{\ell - \lambda}{4} (p_1^2 + p^{\prime \, 2}_1 - k_1^2)
 + (\ell - \lambda) p_0 p^\prime_0
 + \frac{\ell - \lambda}{2} p_1  p^{\prime}_1 +\nonumber\\
&& ~~~~~~~~ ~~~~
+ \frac{\ell + \lambda}{2} (p_0^2 + p^{\prime \, 2}_0 + k_0^2)
 + \frac{\ell + \lambda}{4} (p_1^2 + p^{\prime \, 2}_1 + k_1^2)
 + (\ell + \lambda) (p_0 p^\prime_0 + p_0 k_0 + p^\prime_0 k_0)
 + \frac{\ell + \lambda}{2} (p_1 p^\prime_1 + p_1 k_1 + p^\prime_1 k_1) = \nonumber\\
&& ~~~~~  =  p_0+ p^{\prime}_0+k_0
+ \frac{\ell - \lambda}{2} (p_0 + p^{\prime}_0- k_0) (p_0 + p^{\prime}_0 + k_0)
 + \frac{\ell - \lambda}{4} (p_1 + p^{\prime}_1- k_1) (p_1 + p^{\prime}_1 + k_1)+\nonumber\\
&& ~~~~~~~~ ~~~~
+ \frac{\ell + \lambda}{2} (p_0 + p^{\prime}_0+ k_0)^2
 + \frac{\ell + \lambda}{4} (p_1 + p^{\prime}_1+ k_1)^2 = 0
 \label{covathreeunoHHppk}
\end{eqnarray}
Having done this it is not hard to adapt the results I obtained
for $p \oplus_{\ell} p^\prime \oplus_{\ell , \lambda} k=0$ to the
only slightly different case of $p \oplus_{\ell , \lambda} k \oplus_{\lambda} k'=0$.
In order to establish also the covariance of $p \oplus_{\ell , \lambda} k \oplus_{\lambda} k'=0$
let me start again with its $0$-th component, for which I easily find a satisfactory
 result:
\begin{eqnarray}
&& [N_{[p]}+N_{[k]} +N_{[k^\prime]} + \frac{\ell + \lambda}{2} p_0  ( N_{[k]} +  N_{[k^\prime]})
+ \lambda  k_0 N_{[k^\prime]},
p_0  + k_0 + k^\prime_0]
= p_1+k_1 +k^\prime_1
+ \frac{\ell + \lambda}{2} p_0  (k_1 + k^\prime_1) + \lambda  k_0 k^\prime_1  =\nonumber\\
&& ~~~~~~~~~~~~~~~~~  =(p \oplus_{\ell , \lambda} k \oplus_{\lambda} k')_1= 0
\label{covathreezeroHHpkk}
\end{eqnarray}
And equally satisfactory (but again slightly more tedious) is the corresponding
 verification
for the $1$-component of $p \oplus_{\ell , \lambda} k \oplus_{\lambda} k'=0$
which progresses as follows:
\begin{eqnarray}
&& [N_{[p]}+N_{[k]} +N_{[k^\prime]} + \frac{\ell + \lambda}{2} p_0  ( N_{[k]} +  N_{[k^\prime]})
+ \lambda  k_0 N_{[k^\prime]},
p_1+k_1 + k^\prime_1
+ \frac{\ell + \lambda}{2} p_0  (k_1 + k^\prime_1) + \lambda  k_0 k^\prime_1 ] =\nonumber\\
&& ~~~~~~~~ = p_0+\ell p_0^2 + \frac{\ell}{2} p_1^2
+ k_0+\lambda k_0^2  + \frac{\lambda}{2} k_1^2
+ k^\prime_0+\lambda k^{\prime \, 2}_0 + \frac{\lambda}{2} k^{\prime \, 2}_1 +\nonumber\\
&& ~~~~~~~~ ~~~~
+ \frac{\ell + \lambda}{2} p_1  (k_1 + k^\prime_1) + (\ell + \lambda) p_0 ( k_0+ k^\prime_0)
+ \lambda k_1 k^\prime_1 + 2 \lambda k_0 k^\prime_0 =
 \nonumber\\
&& ~~~~~ = p_0 + k_0  + p^\prime_0
 + \frac{\lambda - \ell }{2} (k_0^2 + k^{\prime \, 2}_0 - p_0^2)
 + \frac{\lambda - \ell }{4} (k_1^2 + k^{\prime \, 2}_1 - p_1^2)
 + ( \lambda - \ell) k_0 k^\prime_0
 + \frac{ \lambda - \ell}{2} k_1  k^{\prime}_1 +\nonumber\\
&& ~~~~~~~~ ~~~~
+ \frac{\ell + \lambda}{2} (p_0^2  + k_0^2 + k^{\prime \, 2}_0)
 + \frac{\ell + \lambda}{4} (p_1^2  + k_1^2 + k^{\prime \, 2}_1)
 + (\ell + \lambda) (p_0 k_0 + p_0 k^\prime_0 + k^\prime_0 k_0)
 + \frac{\ell + \lambda}{2} (p_1 k_1 + p_1 k^\prime_1 +  k^\prime_1 k_1) = \nonumber\\
&& ~~~~~  =  p_0+k_0+ p^{\prime}_0
+ \frac{\lambda - \ell  }{2} (k_0 + k^{\prime}_0- p_0) (k_0 + k^{\prime}_0 + p_0)
 + \frac{\lambda - \ell}{4} (k_1 + k^{\prime}_1- p_1) (k_1 + k^{\prime}_1 + p_1)+\nonumber\\
&& ~~~~~~~~ ~~~~
+ \frac{\ell + \lambda}{2} (p_0 + k_0 + k^{\prime}_0)^2
 + \frac{\ell + \lambda}{4} (p_1 + k_1+ k^{\prime}_1)^2 = 0
 \label{covathreeunoHHpkk}
\end{eqnarray}

So I did manage to establish that it is possible to have meaningful (interacting)
theories that are fully relativistic in spite of allowing for the
coexistence of a type of particle whose relativistic properties
are governed by a DSR $\ell$-deformation of Lorentz symmetry
and of a type of particle whose relativistic properties
are governed by a DSR $\lambda$-deformation of Lorentz symmetry.

\section{Composite particles and potential implications for macroscopic bodies}\label{minibasketsec}
The main result I am here announcing is contained in the previous two sections.
In the remainder of this manuscript my only objective is to show
that the new class of DSR-relativistic theories introduced in the previous two sections
may have application in quite a few different physical pictures:\\
{\bf (i)} It could evidently be used to describe pictures in which different ``elementary/fundamental"
particles have different relativistic properties.\\
{\bf (ii)} It could also be used to describe pictures in which all ``elementary/fundamental"
particles have the same DSR-deformed relativistic properties, but ``composite microscopic particles",
such as atoms
(because of the known mechanisms mentioned in Section I) have different relativistic
properties, with weaker deformation of special-relativistic properties than
the fundamental particles that compose them.\\
{\bf (ii)} And perhaps it could also be used to describe pictures in which microscopic
particles have DSR-deformed relativistic properties, but macroscopic bodies
(again because of the known mechanisms mentioned in Section I) have
ordinary special-relativistic properties.

For cases {\bf (i)} and {\bf (ii)}, assuming indeed $|\ell|^{-1}$ is of the order of the
Planck scale
the new class of DSR-relativistic theories introduced in the previous two sections
certainly provides plausible physical pictures, since for microscopic particles (even for atoms)
the Planck scale is a gigantic scale and all effects of DSR deformation amount
to small corrections.

The physical picture of case {\bf (iii)} instead does not look too promising:
even for this case {\bf (iii)} the fact that I have shown here how
different relativistic properties can coexist is a significant step forward,
but for macroscopic bodies the Planck scale is actually a small energy scale
and there is therefore the risk of predicting hugely unrealistic effects. It is
also for this reason that so far the most popular way to handle macroscopic bodies
in DSR research has been (see, {\it e.g.}, Refs.~\cite{dsrPOLAND2001,soccerball})
the one of renouncing to the introduction of direct interactions between
macroscopic and microscopic particles: the interaction between a macroscopic
particle could be of course also described in terms of the microscopic interactions
involving the constituents of the macroscopic body.\\
It may well be the case that in spite of the option I managed to produce in
the previous two sections one should still proceed in this way. But it is
no longer so obvious that this should be the case:
in this section I am going to report a simple derivation which provides
encouragement for the possibility of allowing direct interactions
between microscopic particles and macroscopic bodies (without needing
to decompose such interactions in terms of the microscopic
interactions between the microscopic particle and the constituents
of the macroscopic body).

The simple derivation I am reporting does not establish anything of general
validity, but it does show that at least some of the derivations which
could turn pathological with a DSR description of macroscopic bodies
actually do not.

This simple derivation concerns an elastic collision between
 an elementary particle and a macroscopic body
$$e^- + X \rightarrow e^- + X$$
where $e^-$ stands for any elementary particle ({\it e.g.} an electron)
governed by a DSR-relativistic description of the type in
Sec.~\ref{kappauno}
\begin{equation}
 p_0= p + \frac{m^2}{2p} - \frac{\ell}{2}  p^2
\label{metricMACRO}
\end{equation}
while $X$ is a macroscopic body of large mass $M$ ($M> |\ell|^{-1}$)
in a nonrelativistic regime but possibly with large spatial momentum $k$
($ |\ell|^{-1} < k \ll M$)
\begin{equation}
 k_0= M + \frac{k^2}{2M}
\label{metricHHk}
\end{equation}
Since I am considering $M> |\ell|^{-1}$ (and $k > |\ell|^{-1}$) one might fear that this elastic scattering
might give pathological results, such as a huge transfer of momentum (of order, {\it e.g.} $\ell k$)
from the macroscopic body to the micro particle. But this is not what the formalization
I developed in the previous two sections predicts.

In seeing this the nontrivial point of the derivation is of course the ``mixing composition law" which
I introduced in Sec.~\ref{hopfliesec}. The process I am here considering has two incoming
and two outgoing particles, so it must be written in terms of two antipodes
(see Ref.~\cite{goldenrule} and references therein)
\begin{equation}
(\ominus_{\ell} p^\prime) \oplus_{\ell} p \oplus_{\ell \star}
 (\ominus_{\star} k^\prime) \oplus_{\star} k = 0~.
\label{connectionMACRO}
\end{equation}
Since $\ominus_{\star}$ is the undeformed composition law
one has that $(\ominus_{\star} k)_\mu = -k_\mu$,
whereas for the $\oplus_{\ell}$ composition law the antipode is such that
still $(\ominus_\ell p)_0 = - p_0$ but $(\ominus_\ell p)_j = - p_j + \ell p_0 p_j$
(in fact $[(\ominus_\ell p) \oplus_{\ell} p]_j = - p_j + \ell p_0 p_j + p_j +\ell (-p_0)p_j=0$).

So for the process I am considering one has
\begin{equation}
- p^\prime_1 + \ell p^\prime_0 p^\prime_1 + p_1 - \ell p^\prime_0 p_1
- k^\prime_1 + k_1 +\frac{\ell}{2} ( p_0 - p^\prime_0) (k_1 - k^\prime_1) = 0~.
\label{connectionMACROmom}
\end{equation}
and
\begin{equation}
- p^\prime_0 + p_0 - k^\prime_0 + k_0 = 0~.
\label{connectionMACROene}
\end{equation}

From (\ref{connectionMACROmom})
one finds that
\begin{equation}
 p^\prime_1 - p_1 \simeq k_1 - k^\prime_1
 +\frac{\ell}{2} ( p_0 + p^\prime_0) (p^\prime_1 - p_1) \simeq
 k_1 - k^\prime_1
 +\ell  p_1 (p^\prime_1 - p_1)~.
\label{connectionMACROmomSH}
\end{equation}
where on the right-hand side, consistently with the fact that I am working throughout
at leading order, I used zero-th order properties in a reexpression of the first-order term.

Then from (\ref{connectionMACROene}) one has that
\begin{equation}
p^\prime_1 + \frac{m^2}{2 p^\prime_1} - \frac{\ell}{2} p^{\prime \, 2}_1 +M+ \frac{k^{\prime \, 2}_1}{2 M}
 = p_1 + \frac{m^2}{2 p_1} - \frac{\ell}{2} p^{2}_1 +M+ \frac{k^{2}_1}{2 M}
 ~,
\label{connectionMACROeneSH1}
\end{equation}
which gives
\begin{equation}
p^\prime_1 - p_1 =\frac{1}{2 M} (k^{2}_1 - k^{\prime \, 2}_1)
+ \frac{m^2}{2 p_1} - \frac{m^2}{2 p^\prime_1} - \frac{\ell}{2} (p^{2}_1 - p^{\prime \, 2}_1)
  \simeq \frac{1}{2 M} (k^{2}_1 - k^{\prime \, 2}_1)
+ \frac{m^2}{2 p_1} - \frac{m^2}{2 p^\prime_1}
 ~,
\label{connectionMACROeneSH2}
\end{equation}
where on the right-hand side I used again
zero-th order properties in a reexpression of the first-order term.

Combining (\ref{connectionMACROmomSH}) and (\ref{connectionMACROeneSH2}) one sees that,
at least in the specific context of an elastic collision between a DSR-deformed micro particle
and special-relativistic macroscopic body no pathology arises:
neither in  (\ref{connectionMACROmomSH}) nor in (\ref{connectionMACROeneSH2})
one finds pathological correction terms of the type $\ell k$ or $\ell M$.\\
If it turned out to be possible to extend this observation to a wider class
of phenomena we might also have
an exciting opportunity for the description of ``observers"
in DSR-relativistic theories of this sort. In any relativistic theory an
observer is to be identified with a macroscopic device (or, idealizing,
a network of such devices),
and so far studies of this
type of on-shell-relation-centered DSR scenarios have kept such observers in a sort
of ``limbo", protected artificially (by hand) from the implications of the
deformation of relativistic symmetries.
If it turned out to be possible to generalize the observation I reported in this
section those artificial abstractions could be eliminated in favor of a
more satisfactory picture of DSR observers.

\section{Potential implications for OPERA-anomaly-type phenomenology}\label{operasec}
Having discussed a possible implication of the results I reported in Secs.~\ref{hopfliesec}
and \ref{hopfhopfsec} for the inclusion of special-relativistic macroscopic bodies
in an otherwise DSR-relativistic framework, in this section I go to the opposite
extreme of the range of possible applications of the
results I reported in Secs.~\ref{hopfliesec}
and \ref{hopfhopfsec}, by considering the possibility that different microscopic (possibly ``fundamental")
particles be governed by different DSR-relativistic properties.

As I already stressed above (this was characterized as case {\bf (i)}  in the previous section)
this possible application is clearly viable if one
assumes indeed that $|\ell|^{-1}$ is of the order of the
Planck scale (or some other ultralarge momentum scale),
since for microscopic particles (even for atoms)
the Planck scale is gigantic and all effects of DSR deformation amount
to small corrections.

While this is evident, it is nonetheless valuable for me to stress
that, in spite of the apparently ``invasive" prescription of different
DSR-relativistic properties for different types of particles, the deformation
is still a very smooth deformation of special-relativistic symmetries.
I shall do this by showing that some pathologies that are expected
in cases where different particles have different relativistic properties
(if this is the result of a full breakdown of relativistic symmetries, with emergence
of a preferred frame), are not present when this feature is introduced in the
DSR-compatible manner I proposed in this manuscript.

In order to give some tangibility to my discussion (well, some tentative tangibility)
I shall take as illustrative example the type of phenomenology of departures
from Lorentz symmetry that was recently motivated by the fact that the
OPERA collaboration reported~\cite{opera} tentative evidence of superluminality for neutrinos.

The possibility that superluminal particles might well be describable in DSR-compatible fashion
(without the introduction of a preferred frame) has already been raised
in a few OPERA-motivated papers
(see, {\it e.g.}, Refs.~\cite{whataboutopera,operaDSR,fransDSROPERA,synchroDSR,yiDSROPERA,dimitriDSROPERA}).

The fact that particles with the DSR-relativistic properties here described
in Sec.~\ref{kappauno} (and \ref{hopfliesec}
and \ref{hopfhopfsec}) are ``superluminal" for negative $\ell$ ({\it i.e.} $\ell/|\ell| = -1$)
has been established in several previous
studies~\cite{gacMandaniciDANDREA,bob,leeINERTIALlimit,k-bob,anatomy}: when these particles have spatial
momentum $p$ such that $\ell p > m^2/p^2$ their speed is higher than the speed
of low-energy photons, and in that sense they indeed are ``superluminal".

But, while previous quantum-gravity-inspired studies of DSR-relativistic theories
usually assumed ``universality",
it appears that any attempt to do OPERA-inspired phenomenology should require
a non-universal picture: for photons of, say, about 20GeV the agreement of the
speed law with the special-relativistic prescription is confirmed at the
level of 1 part in $10^{18}$ (see, {\it e.g.}, Refs.~\cite{hessVGAMMA,fermiSCIENCE,FERMI090510}),
whereas taking the OPERA result at face value
one should assume that $\sim 20$-GeV neutrinos experience departures from
the special-relativistic speed law at the level of a few parts in $10^5$.

Ref.~\cite{operaDSR} observed that one might well have a universal DSR deformation
and yet have ``effectively particle-dependent effects": the illustrative example adopted in
Ref.~\cite{operaDSR} is centered on an on-shell relation of the type
\begin{equation}
 p_0^2 = p^2 + m^2 + 2 \ell \frac{p_0^2 p^2}{m} ~,
 \label{dispGAC}
 \end{equation}
universally for all particles\footnote{Massless particles should anyway be excluded
in Eq.~(\ref{dispGAC}).
A more sophisticated way to achieve similar goals
would be to adopt a dispersion relation of the type
$$p_0^2 = p^2 + m^2 + \ell^2 \frac{ p^6}{p_0^2-p^2}$$}, but yet effectively attributing stronger
departures from
special relativity to lighter particles, because of the explicit dependence on the mass
(a relativistic invariant) in the correction term.

This idea of a universal DSR deformation with ``effectively particle-dependent effects"
may well be valuable in OPERA-inspired phenomenology.
However, from this perspective,
I have provided in this manuscript an interesting alternative
by showing that it is possible to have a consistent DSR-relativistic
framework even endowing different types of particles with genuinely different DSR-relativistic
properties. It is now possible to contemplate genuinely ``non-universal" DSR-relativistic pictures.

In light of this it is an amusing exercise, to which I devote the remainder of
this section, to take indeed negative $\ell$
and assume for the purposes of the exercise that $|\ell|^{-1}$ might be much smaller
than the Planck scale, but still much higher than 20 GeV. This will allow me
to contemplate some of the issues that have taken center stage
in the OPERA-related literature, as possible pathologies of superluminal neutrinos,
and show that because of the smoothness of the DSR deformations (even the ones I here
introduced, with a dependence on the type of particle) these concerns
do not apply to the DSR formulation I here introduced as first illustrative example
of ''non-universal" DSR-relativistic framework.

My OPERA-inspired exercise will be conducted (for the sake of the argument)
assuming that only neutrinos are governed by the DSR picture of Sec.~\ref{kappauno},
with
\begin{equation}
 p_0= p + \frac{m_\nu^2}{2p} - \frac{\ell}{2}  p^2~,
\label{metricCHEREp}
\end{equation}
while all other microscopic particles are also
governed by the DSR picture of Sec.~\ref{kappauno} but with a weaker
deformation
\begin{equation}
 k_0= k + \frac{m_x^2}{2k} - \frac{\lambda}{2}  k^2
\label{metricCHEREk}
\end{equation}
with $\lambda$ also negative and $0 \leq |\lambda| \leq |\ell|$
(importantly including the case $\lambda=0$, $\ell \neq 0$ and the case $\lambda=\ell$
as limiting cases).

\subsection{Absence of Cherenkov-like $\nu \rightarrow \nu + X$ processes}\label{noglashowsec}
Among the possible ``pathologies" for superluminal neutrinos
which have been of interest in OPERA-inspired phenomenology much attention
has been devoted~\cite{glashowOPERA} (also see, {\it e.g.}, Refs.~\cite{liberatiOPERA,joseOPERA})
 to {\underline{in-vacuo}}
 Cherenkov-like processes $\nu \rightarrow \nu + X$ (where $X$ may also be, {\it e.g.}
 an electron-positron pair).
In-vacuo Cherenkov-like processes are forbidden in special relativity, but in theories
with superluminal neutrinos, {\underline{if they violate/break relativistic invariance}}
(with associated emergence of a preferred ``ether" frame),
they can be allowed if the energy of the incoming neutrino
is above a certain threshold value.

Ref.~\cite{operaDSR} (on the basis of results already discussed, for other purposes,
in Refs.~\cite{gacnewjourn,sethmajor})
already observed that in standard ``universal" DSR-relativistic descriptions
of neutrino superluminality, which are fully relativistic and do not have a preferred frame,
such in-vacuo Cherenkov-like processes remain forbidden, in spite of the superluminality
of the neutrinos.

The notion of ``non-universal" DSR-relativistic framework,
which I here introduced, also fully preserves the relativity of inertial frames,
so in-vacuo Cherenkov-like processes must also be forbidden~\cite{goldenrule,gacnewjourn}
in this new class of relativistic theories. But it is still valuable to see this
in an explicit calculation.
So let me analyze the relativistic kinematics of $$\nu \rightarrow \nu + X$$
in the novel setup of Sec.~\ref{hopfhopfsec}, assuming indeed, as I announced for this
section, that neutrinos are affected by deformation with parameter $\ell$
while all other particles are affected by deformation with parameter $\lambda$,
with $0 \leq |\lambda| \leq |\ell|$.

Within the framework here introduced in Sec.~\ref{hopfhopfsec}
such a process should be described in terms of the ``mixing conservation law"
\begin{equation}
(\ominus_{\ell} p^\prime) \oplus_{\ell} p \oplus_{\ell \lambda} k = 0~.
\label{connectionCHERE}
\end{equation}
where $p$ is the (four-)momentum of the outgoing neutrino, $k$ is the momentum
of the outgoing $X$, and $\ominus_{\ell} p^\prime$ is the $\oplus_{\ell}$-antipode of the
momentum of the incoming neutrino.

In light of the properties specified for $\oplus_{\ell}$
in Secs.~\ref{kappauno}, \ref{hopfliesec}, \ref{hopfhopfsec},
one finds that the $0$-component of this conservation law is
\begin{equation}
0 = - p^\prime_0 + p_0 + k_0 =
- p^\prime_1 - \frac{m^2}{2 p^\prime_1} + \frac{\ell}{2} p^{\prime \, 2}_1
+ p_L + \frac{m^2 + p_T^2}{2 p_L} - \frac{\ell}{2} p^{2}_L
+ k_L + \frac{m_x^2 + p_T^2}{2 k_L} - \frac{\ell}{2} k^{2}_L
~.
\label{connectionCHEREene}
\end{equation}
where I also used (\ref{metricCHEREp}) and (\ref{metricCHEREk})
with $m$ the mass of the incoming and outgoing neutrino,
and $m_x$ the rest energy of $X$.
Also notice that I am assuming that the incoming neutrino is ultrarelativistic
(in particular $p \gg m_x$) and the momenta transverse to the direction of the
incoming neutrino are small: along that transverse direction both the outgoing
neutrino and the outgoing $X$ have transverse momenta of  roughly the same magnitude,
here denoted with $p_T$,
which is very small compared to the momenta these outgoing particles have
along the direction of the incoming neutrino.

For the momenta (with index $_L$) along that direction of the incoming neutrino
one obtains from (\ref{connectionCHERE})
\begin{equation}
- p^\prime + \ell p^\prime_0 p^\prime + p_L - \ell p^\prime_0 p_L
+ k_L +\frac{\ell + \lambda}{2} ( p_0 - p^\prime_0) k_L = 0~.
\label{connectionCHEREmom}
\end{equation}
{\it i.e.}
\begin{equation}
 p^\prime = p_L + k_L
 + \ell  p^\prime_0 k_L - \frac{\ell + \lambda}{2} k_0 k_L
\label{connectionCHEREmomSH}
\end{equation}
where I used again in the leading-order correction properties established at 0-th order.\\
Combining (\ref{connectionCHEREene}) with (\ref{connectionCHEREmomSH}) one
obtains
\begin{equation}
\frac{p_T^2}{2 p_L} +  \frac{ p_T^2}{2 k_L} ~ = ~
 \frac{m^2}{2 p^\prime_1} - \frac{m^2}{2 p_L} - \frac{m_x^2}{2 k_L}
+ \ell  p^\prime_0 k_L - \frac{\ell + \lambda}{2} k_0 k_L
  - \frac{\ell}{2} p^{\prime \, 2}_1
+ \frac{\ell}{2} p^{2}_L
+  \frac{\ell}{2} k^{2}_L ~ \simeq ~
 \frac{m^2}{2 p^\prime_1} - \frac{m^2}{2 p_L} - \frac{m_x^2}{2 k_L}
 ~,
\label{connectionCHEREeneSH2}
\end{equation}
where on the right-hand side I observed that
$$ \ell  p^\prime_0 k_L - \frac{\ell + \lambda}{2} k_0 k_L
  - \frac{\ell}{2} p^{\prime \, 2}_1
+ \frac{\ell}{2} p^{2}_L
+  \frac{\ell}{2} k^{2}_L \simeq 0$$
using properties valid at 0-th order in rearranging this leading-order correction.

Eq.~(\ref{connectionCHEREeneSH2}) is the main result of this subsection.
For scenarios with neutrino superluminality and breakdown of Lorentz symmetry, with emergence
of a preferred frame, the key observation concerning
in-vacuo Cherenkov-like processes is that the analog of Eq.~(\ref{connectionCHEREeneSH2})
contains strong corrections~\cite{glashowOPERA} from Lorentz-symmetry-breaking terms such that
then real values of $p_T$ are allowed. One way to see that in-vacuo Cherenkov-like processes
are forbidden in special relativity is through the fact that it would require imaginary
values of $p_T$. And in my DSR-relativistic analysis I found, as codified in
Eq.~(\ref{connectionCHEREeneSH2}), that independently of the value of the momentum
of the incoming neutrino the $p_T$ is necessarily imaginary;
in-vacuo Cherenkov-like processes are indeed still forbidden in my DSR-relativistic
framework.

\subsection{Absence of anomalies for $\pi \rightarrow \mu + \nu$}\label{nonussisec}
Another ``pathology" for superluminal neutrinos
which has been of interest in OPERA-inspired phenomenology
concerns the pion-decay channel  $\pi \rightarrow \mu + \nu$, for which, in theories
with superluminal neutrinos, {\underline{if they violate/break relativistic invariance}}
(with associated emergence of a preferred ``ether" frame),
one finds that for high-energy pions the  muon/muon-neutrino phase space
available for the decay is severely reduced~\cite{nussiOPERA,gonzaOPERA,Bi}.

Also with respect to this other concern for OPERA-inspired phenomenology
the notion of ``non-universal" DSR-relativistic framework,
which I here introduced, turns out to be immune,
mainly as a result of the
fact that it fully preserves the relativity of inertial frames.
To see this I shall analyze the relativistic kinematics of the process
$$\pi \rightarrow \mu + \nu $$
in the novel setup of Sec.~\ref{hopfhopfsec}, assuming indeed, as I announced for this
section, that neutrinos are affected by deformation with parameter $\ell$
while all other particles are affected by deformation with parameter $\lambda$,
with $0 \leq |\lambda| \leq |\ell|$, so that in particular
\begin{equation}
 p_{\nu 0}= p_{\nu L} + \frac{m_\nu^2+p_T^2}{2p_{\nu L}} - \frac{\ell}{2}  p_{\nu L}^2
\label{metricPIDECnu}
\end{equation}
\begin{equation}
 k_{\pi 0}= k_\pi + \frac{m_\pi^2}{2k_\pi} - \frac{\lambda}{2}  k_\pi^2
\label{metricPIDECpi}
\end{equation}
\begin{equation}
 k_{\mu 0}= k_{\mu L} + \frac{m_\mu^2+p_T^2}{2k_{\mu L}} - \frac{\lambda}{2}  k_{\mu L}^2
\label{metricPIDECmu}
\end{equation}
I am evidently again focusing on the case that the incoming particle (this time the pion)
is ultrarelativistic
($p_\pi \gg m_\pi$) and the momenta transverse to the direction of the
incoming pion are small: along that transverse direction both the outgoing
neutrino and the outgoing muon have transverse momenta of  roughly the same magnitude,
here denoted again with $p_T$,
which is very small compared to the momenta these outgoing particles have
along the direction of the incoming pion.

On the basis of the findings here reported in Sec.~\ref{hopfhopfsec}
the process $\pi \rightarrow \mu + \nu$
should be described in terms of the ``mixing conservation law"
\begin{equation}
p_\nu \oplus_{\ell \lambda}  [k_\mu \oplus_{\lambda} (\ominus_{\lambda} k_\pi)] = 0~.
\label{connectionPIDEC}
\end{equation}
where  $\ominus_{\lambda} k_\pi$ is the $\oplus_{\lambda}$-antipode of the
momentum of the incoming pion.

For the $0$ component this gives
\begin{equation}
0 = - k_{\pi \, 0} + p_{\nu \, 0} + k_{\mu \, 0} =
- k_{\pi} - \frac{m_\pi^2}{2 k_\pi} +  \frac{\lambda}{2} k^{2}_{\pi}
+ p_{\nu L} + \frac{m_\nu^2+p_T^2}{2 p_{\nu L}}
- \frac{\ell}{2} p^{2}_{\nu L}
+k_{\mu L} + \frac{m_\mu^2+p_T^2}{2 k_{\mu L}}
-  \frac{\lambda}{2} k^{2}_{\mu L}
~.
\label{connectionPIDECene}
\end{equation}
where I also used (\ref{metricPIDECnu}), (\ref{metricPIDECpi}) and (\ref{metricPIDECmu}).

For the momenta (with index $_L$) along the direction of the incoming pion
one obtains from (\ref{connectionPIDEC}):
\begin{equation}
 p_{\nu L} + k_{\mu L} - k_{\pi} + \lambda k_{\pi 0} k_{\pi} - \lambda k_{\mu 0} k_{\pi}
 +\frac{\ell + \lambda}{2}  p_{\nu 0} (k_{\mu L} - k_{\pi}) = 0~.
\label{connectionPIDECmom}
\end{equation}
{\it i.e.}
\begin{equation}
 k_\pi = p_{\nu L} + k_{\mu L} +
 + \ell  p_{\nu 0} k_\pi  - \frac{\ell + \lambda}{2} p_{\nu 0} p_{\nu L}
\label{connectionDECmomSH}
\end{equation}
where again on the right-hand side I used properties of the 0-th order result in
rearranging the leading-order correction.\\
Combining (\ref{connectionPIDECene}) with (\ref{connectionDECmomSH}) one
obtains
\begin{equation}
\frac{p_T^2}{2 p_{\nu L}} +  \frac{ p_T^2}{2 k_{\mu L}} ~ = ~
 \frac{m_\pi^2}{2 k_\pi} - \frac{m_\nu^2}{2 p_{\nu L}} - \frac{m_\mu^2}{2 k_{\mu L}}
+ \ell  p_{\nu 0}  k_\pi  - \frac{\ell + \lambda}{2} p_{\nu 0} p_{\nu L}
+ \frac{\ell}{2} p^{2}_{\nu L}
+  \frac{\lambda}{2} k^{2}_{\mu L}
-  \frac{\lambda}{2} k^{2}_{\pi} ~ \simeq ~
\frac{m_\pi^2}{2 k_\pi} - \frac{m_\nu^2}{2 p_{\nu L}} - \frac{m_\mu^2}{2 k_{\mu L}}
 ~,
\label{connectionPIDECeneSH2}
\end{equation}
where on the right-hand side
I observed that
$$ \ell  p_{\nu 0}  k_\pi  - \frac{\ell + \lambda}{2} p_{\nu 0} p_{\nu L}
+ \frac{\ell}{2} p^{2}_{\nu L}
+  \frac{\lambda}{2} k^{2}_{\mu L}
-  \frac{\lambda}{2} k^{2}_{\pi} \simeq 0$$
using properties valid at 0-th order in rearranging this leading-order correction.

Eq.~(\ref{connectionPIDECeneSH2}) is the main result of this subsection.
For scenarios with neutrino superluminality and breakdown of Lorentz symmetry, with emergence
of a preferred frame, the key observation concerning $\pi \rightarrow \mu + \nu $
 is that the analog of Eq.~(\ref{connectionPIDECeneSH2})
contains strong corrections~\cite{nussiOPERA,gonzaOPERA,Bi}
from Lorentz-symmetry-breaking terms, such that
the combinations of $p_{\nu L}$ and $k_{\mu L}$ which satisfy the requirement $p_T^2 >0$
only amount to a very small phase space.
Instead in my DSR-relativistic analysis I found
Eq.~(\ref{connectionPIDECeneSH2}), in which all correction terms canceled each other out,
 so that the phase space available for the decay in the DSR case is identical to the
 phase space available for the decay in the standard special-relativistic case.\\
 I should stress that this result holds in a leading-order analysis, and it is therefore
 reliable only for $p_\pi \ll |\ell|^{-1}$ (which however is the case of interest for
 Refs.~\cite{nussiOPERA,gonzaOPERA,Bi}). For $p_\pi \sim |\ell|^{-1}$ one might
 have sizeable modifications of the phase space even in a DSR case.

\subsection{Back to quantum gravity and the Planck scale}
The observations reported in the previous two subsections may be used in attempts
of interpreting the OPERA anomaly as an actual manifestation of physics
beyond the reach of special relativity. But chances are the OPERA anomaly will
be eventually found to be described by much less exotic physics. One should therefore
expect that the most promising applications of the results in the previous two subsections
will be in the context where DSR-deformations of Lorentz symmetry were first
conceived, studies of the quantum-gravity problem, for which such deformations
would not be surprising, in the case with deformation scale roughly of the order of
the Planck scale.

There were  some studies (see, {\it e.g.}, Refs.~\cite{urrutiaPRL,alfaroNU2007,ellisONLYPHOTON})
proposing mechanisms by which quantum-gravity/quantum-spacetime effects could
produce departures from Lorentz symmetry of different magnitude for
different particles. The results I reported in this manuscript show that
such scenarios do not necessarily have to ``break" Lorentz invariance, producing
a preferred ``ether" frame. One could attempt to discuss such scenarios
in terms of the particle-type-dependent DSR-deformations of Lorentz symmetry
I here proposed. And then the results reported in the previous two subsections
could be valuable assets for the corresponding ``Planck-scale phenomenology".

From that perspective I should stress
that, as the careful reader can easily verify,
the point made in Subsec.~\ref{noglashowsec} for the
process $\nu \rightarrow \nu + X$ can be generalized to all Cherenkov-like processes $A \rightarrow A + X$.
And similarly
the point made in Subsec.~\ref{nonussisec} for the
process $\pi \rightarrow \mu + \nu$
 can be generalized to all decay processes $A \rightarrow B + C$.

\section{Aside on a ``$\kappa\kappa$-Minkowski spacetime" and algebras
with ``mixing co-products"}\label{mixingcoprodsec}

\subsection{Contemplating algebras
with ``mixing co-products"}
I have here introduced a new class of (DSR-)relativistic theories, focusing in this
first study on kinematics. I expect that in order to achieve a full empowerment
of this new class of theories it will be necessary to identify a symmetry-algebra
counter-part to the novel type of relativistic kinematics I here proposed.
This balance is rather visible in special relativity, whose full understanding
requires combining the Poincar\'e symmetry algebra
and Einstein kinematics.\\
And this ``balance of powers" appears to be preserved also in
the illustrative example of ``universal"
DSR-relativistic theory which I here took as starting point, in Sec.~\ref{kappauno}:
for the specific example of DSR deformation of relativistic kinematics
here reviewed in Sec.~\ref{kappauno}
one can find numerous points of contact with the structure of the $\kappa$-Poincar\'e Hopf algebra~\cite{majidruegg,kpoinap}.
In particular, the most crucial aspect of that construction, the composition law
\begin{equation}
(p \oplus_\ell p^\prime)_1 = p_1 + p^\prime_1 + \ell p_0 p^\prime_1  ~,~~~
(p \oplus_\ell p^\prime)_0 = p_0 + p^\prime_0~,
\label{connectionSECKAPPAUNO}
\end{equation}
can be placed in correspondence with the law of ``co-product"
that characterizes the $\kappa$-Poincar\'e Hopf algebra in the Majid-Ruegg basis~\cite{majidruegg},
which in leading order reads
\begin{equation}
\Delta(P_1) \simeq P_1 \otimes 1 + 1 \otimes P_1 +  \frac{1}{\kappa} P_0 \otimes P_1   ~,~~~
\Delta(P_0) = P_0 \otimes 1 + 1 \otimes P_0   ~.
\label{coprodSECKAPPAUNO}
\end{equation}

In order to establish a similar connection between relativistic kinematics and symmetry algebras
for the novel class of relativistic theories I here proposed in
Secs.~\ref{hopfliesec} and \ref{hopfhopfsec} it would seem necessary to introduce
on the algebra side enough structure to accommodate at least 3 laws of coproduct:
two different laws of the type (\ref{coprodSECKAPPAUNO})
\begin{eqnarray}
\Delta_\ell(P_1)  \simeq P_1 \otimes 1 + 1 \otimes P_1 +  \ell P_0 \otimes P_1   &~,~~~&
\Delta_\ell(P_0) = P_0 \otimes 1 + 1 \otimes P_0   ~, \nonumber\\
\Delta_\lambda(P_1)  \simeq P_1 \otimes 1 + 1 \otimes P_1 +  \lambda P_0 \otimes P_1   &~,~~~&
\Delta_\lambda(P_0) = P_0 \otimes 1 + 1 \otimes P_0   ~, \nonumber
\label{twocoprods}
\end{eqnarray}
and a novel ``mixing coproduct" of a type illustrated by
\begin{equation}
\Delta_{\ell \, \lambda}(P_1)  \simeq P_1 \otimes 1 + 1 \otimes P_1
+  \frac{\ell + \lambda}{2} P_0 \otimes P_1   ~,~~~
\Delta_{\ell \, \lambda}(P_0) = P_0 \otimes 1 + 1 \otimes P_0   ~.
\label{mixincoprod}
\end{equation}
I am not aware of any work on symmetry algebras (some sort of ``$\kappa\kappa$-Poincar\'e algebra")
already providing
these structures, but it appears natural to expect
 that such a construction should be possible.

\subsection{Toward a ``$\kappa\kappa$-Minkowski spacetime"}
There is at least one more ingredient
that would be desirable, in addition to looking for a ``symmetry-algebra counterpart",
as an enrichment of  the
non-universal DSR deformations I here introduced focusing temporarily on
aspects pertaining kinematics from a momentum space perspective.
This other desired ingredient is (one form or another of) a ``spacetime picture".\\
I shall not speculate much about this here. I expect it may require
several striking steps of abstraction, which will take time to mature.
Even for universal DSR deformation we are still in the ``digestion process"
for some of the striking new features that the associate spacetime pictures
typically introduce, such as the relativity of spacetime locality~\cite{bob,leeINERTIALlimit}.
We must expect features possibly even more virulent (but again not necessarily
in conflict with established experimental facts) to be typical for
the novel non-universal DSR deformations I am here introducing.

Hoping to offer a useful contribution to the study of these issues
I venture to formulate here only a preliminary speculation.
This concerns the fact that
for the illustrative example of ``universal"
DSR-relativistic theory which I here took as starting point, in Sec.~\ref{kappauno},
there is an established ``formal link" to the so-called ``$\kappa$-Minkowski
spacetime" (see, {\it e.g.},Refs.~\cite{majidruegg,kpoinap,gacAlessandraFrancesco}),
in which the same scale $\kappa$
of (\ref{coprodSECKAPPAUNO}) appears in a form of spacetime noncommutativity
\begin{equation}
[{\hat x}_j,{\hat t}] = i \frac{1}{\kappa} {\hat x}_j~,~~~[{\hat x}_j,{\hat x}_k] =0~.
\label{kappamink}
\end{equation}
The core feature of this formal link is visible already in how
the composition law (\ref{connectiontorsy})
\begin{equation}
(p \oplus_\ell p^\prime)_1 = p_1 + p^\prime_1 + \ell p_0 p^\prime_1
\label{connectiontorsyONE}
\end{equation}
emerges in cases where certain ordering prescriptions are applied
in analyses of $\kappa$-Minkowski noncommutativity, such as in
\begin{equation}
e^{i p_j {\hat x}^j}
e^{i p_0 {\hat t}} e^{i p^\prime_j {\hat x}^j}
= e^{i p_j {\hat x}^j}  e^{i  e^{\ell p_0} p^\prime_j {\hat x}^j}
e^{i p_0 {\hat t}} \simeq
e^{i [p_j + p^\prime_j + \ell p_0 p^\prime_j] {\hat x}^j}
e^{i p_0 {\hat t}}
\label{connectiontorsyONE}
\end{equation}
where I used (\ref{kappamink}) for $\ell \equiv 1/\kappa$.

Inspired by this observation,
one may consider attempting to provide a formal spacetime picture
for a non-universal DSR-relativistic scenario of the type in Sec.~\ref{hopfhopfsec}
by seeking a generalization/deformation of $\kappa$-Minkowski spacetime,
suitable for accommodating the ``mixing composition law" introduced in
Sec.~\ref{hopfhopfsec}, and particularly, from (\ref{connectionHHpk}),
\begin{equation}
(p \oplus_{\ell,\lambda} k)_j = p_j + k_j + \frac{\ell + \lambda}{2} p_0 k_j  ~,~~~
\label{connectionHHpkREMAKE}
\end{equation}
or similarly\footnote{Note that a conservation law of the
type $p_0 + k_0 =0~,~~p_j + k_j + (\ell + \lambda) p_0 k_j/2=0$
is equivalent to $p_0 + k_0 =0~,~~p_j + k_j + \lambda  p_0 k_j/2-\ell k_0 p_j/2=0$
since (for $p_0 + k_0 =0$ and working at leading order) one has that
$$p_j + k_j + \frac{\ell + \lambda}{2} p_0 k_j=0
~~\Rightarrow ~~ (1+\ell k_0/2) [p_j + k_j + \frac{\ell + \lambda}{2} p_0 k_j]=0 ~~ \Leftrightarrow ~~
p_j + k_j + \frac{\lambda}{2} p_0 k_j-\frac{\ell}{2} k_0 p_j=0 $$}
\begin{equation}
(p \oplus_{\ell,\lambda} k)_j = p_j + k_j + \frac{\lambda}{2} p_0 k_j-\frac{\ell}{2} k_0 p_j ~,~~~
\label{connectionHHpkREMAKEbis}
\end{equation}
I feel that one tempting possibility would be the one of contemplating
a notion of quantum spacetime in which the coordinates of different types
of particles have different noncommutativity properties, which taking as starting
point $\kappa$-Minkowski spacetime may lead one to consider a generalization
suitable for being labeled as ``$\kappa\kappa$-Minkowski spacetime".\\
Postponing a more in depth analysis of possible alternatives to future work,
I just want to notice, concerning this $\kappa\kappa$-Minkowski speculation,
that structures such as the one found in (\ref{connectionHHpkREMAKE})
and (\ref{connectionHHpkREMAKEbis}) might be naturally encountered
 in cases where suitable ordering prescriptions are applied
in analyses of a scenario such that my ``$p$-particles" have coordinates
with noncommutativity
\begin{equation}
[{\hat x}_j,{\hat t}] = i \ell {\hat x}_j~,~~~[{\hat x}_j,{\hat x}_k] =0~.
\label{kappaminkB}
\end{equation}
while my ``$k$-particles" have coordinates
with noncommutativity
\begin{equation}
[{\hat x}'_j,{\hat t}] = i \lambda {\hat x}'_j~,~~~[{\hat x}'_j,{\hat x}'_k] =0~.
\label{kappaminkC}
\end{equation}
I conjecture that some of the findings of recent works on
a possible relativity of spacetime locality~\cite{bob,leeINERTIALlimit,prl}
will lead to a conceptualization of spacetime for which it would not be
cumbersome to introduce different types of particles ``sharing the same (relative-locality) spacetime"
and yet with coordinates governed by different laws of noncommutativity.\\
And on the basis of (\ref{kappaminkB}) and  (\ref{kappaminkC}) one could produce
terms of the form $\lambda p_0 k_j$
from applying suitable ordering prescriptions in such a $\kappa\kappa$-Minkowski spacetime,
as shown by
\begin{equation}
e^{i p_j {\hat x}^j}
e^{i p_0 {\hat t}} e^{i k_j {\hat x}^j}
= e^{i p_j {\hat x}^j}  e^{i  e^{\lambda p_0} k_j {\hat x}'^j}
e^{i p_0 {\hat t}} \simeq
e^{i [p_j {\hat x}^j+ (k_j + \lambda p_0 k_j) {\hat x}'^j]}
e^{i p_0 {\hat t}}
\label{connectiontorsyONEbbb}
\end{equation}
and one could produce
terms of the form $\ell k_0 p_j$
from other applications of
suitable ordering prescriptions in
such a $\kappa\kappa$-Minkowski spacetime, as shown by
\begin{equation}
e^{i k_j {\hat x}'^j}
e^{i k_0 {\hat t}} e^{i p_j {\hat x}^j}
= e^{i k_j {\hat x}'^j}  e^{i  e^{\ell k_0} p_j {\hat x}^j}
e^{i p_0 {\hat t}} \simeq
e^{i [p_j {\hat x}^j+ (k_j + \lambda p_0 k_j) {\hat x}'^j]}
e^{i k_0 {\hat t}}
\label{connectiontorsyONEccc}
\end{equation}

\section{Aside beyond leading order}\label{allordersec}
I have worked throughout this manuscript at leading order in the deformation scale.
As stressed above this is the only reasonable choice since in quantum-gravity/Planck-scale
phenomenology it would be already very fortunate to ever uncover leading-order effects
and at least presently going beyond leading order is unjustified.
As I stressed already in Refs.~\cite{dsr1Edsr2,dsrPOLAND2001}, enforcing some sort
of requirement of mathematical consistency beyond leading order is not only inappropriate because
of the limitations of the expected experimental sensitivities, but may also
be inappropriate in light of the complexity of the quantum-gravity problem:
even if DSR-deformations of Lorentz symmetry do end up being
actually relevant for quantum gravity (and of course
this is only a remote hypothesis)
it may well be the case that only their ``leading-order formulation"
makes sense physically. This is because the availability of a
description in terms of DSR deformations
essentially still assumes a rather standard picture of spacetime, novel enough to include
striking new features such as a relativity of spacetime locality~\cite{bob,leeINERTIALlimit,prl},
but still conventional enough to allow a description to a large extent still consistent
with the standard role of spacetime in physics. However, the nature of the quantum-gravity
problem is such that it would not be surprising if, as the characteristic energy scales
get closer to the Planck scale (just as the leading-order analysis starts to be
insufficient), at some point there would be the onset of a completely foreign regime
of the laws of physics, not even affording us the luxury of the abstraction of
an (however exotic) spacetime formulation.

So I do not view the development of DSR pictures beyond leading order as
 an important priority. It is nonetheless conceptually intriguing and provides amusing challenges.
 I shall not dwell much here on this issue, but let me nonetheless at least exhibit, in this section,
 partial results that provide some encouragement for the idea that
 such ``all-order particle-type-dependent DSR-deformations" are indeed
 possible.

The first ingredient I introduce for this purpose is an ``all-order generalization"
of the type of particle described in Sec.~\ref{kappauno}.
This generalization replaces Eqs.~(\ref{metrictorsy})
(\ref{connectiontorsy}), (\ref{booststorsy}), (\ref{boosttwo}) with
\begin{equation}
\cosh (\ell m) = \cosh (\ell p_0) - \frac{\ell^2}{2} e^{-\ell p_0} p_1^2
\label{metrictorsyALL}
\end{equation}
\begin{equation}
(p \oplus_\ell p^\prime)_1 = p_1 + e^{\ell p_0} p^\prime_1  ~,~~~
(p \oplus_\ell p^\prime)_0 = p_0 + p^\prime_0~.
\label{connectiontorsyALL}
\end{equation}
\begin{equation}
[N, p_0] =  p_1 ~,~~~
[N, p_1] = \frac{e^{2 \ell p_0} - 1}{2\ell}  + \frac{\ell}{2} p_1^2~,
\label{booststorsyALL}
\end{equation}
\begin{eqnarray}
N_{[p \oplus_\ell p^{\prime}]} =N_{[p]}+ e^{\ell p_0} N_{[p^{\prime}]}
\label{boosttwoALL}
\end{eqnarray}
These formulas may be viewed as a kinematical counterpart
from some of the structures, including the
mentioned co-products, of
one of the descriptions~\cite{majidruegg} of the
the $\kappa$-Poincar\'e Hopf algebra.

In this section
I refer to particles governed by
Eqs.~(\ref{metrictorsyALL})
(\ref{connectiontorsyALL}), (\ref{booststorsyALL}), (\ref{boosttwoALL})
as ``$p$-particles"
and denote their momenta consistently with $p$ (or $p^{\prime}$ or $p^{\prime\prime}$ ...).

It is easy to see that
Eqs.~(\ref{metrictorsyALL})
(\ref{connectiontorsyALL}), (\ref{booststorsyALL}), (\ref{boosttwoALL})
ensure relativistic consistency for the description of such particles.
In particular,
\begin{equation}
[N,  \cosh (\ell p_0) - \frac{\ell^2}{2} e^{-\ell p_0} p_1^2 ]
 = 0
\label{invariantshelltorsyALL}
\end{equation}
and interactions among ``$p$-particles", with conservation
law $p \oplus_\ell p^\prime \oplus_\ell p^{\prime \prime} = 0$,
evidently admit consistent relativistic description, as shown by
the following two equations:
\begin{eqnarray}
&& [N_{[p]}+ e^{\ell  p_0} N_{[p^\prime]}+ e^{\ell  (p_0+p^\prime_0)} N_{[p^{\prime \prime}]},
p_0 + p^\prime_0 + p^{\prime \prime}_0]
= p_1 + e^{\ell p_0}p^\prime_1 + e^{\ell  (p_0+p^\prime_0)}p^{\prime \prime}_1 =
(p \oplus_{\ell} p^\prime \oplus_{\ell} p^{\prime \prime})_1= 0
\label{covathreezeroHHppkALL}
\end{eqnarray}
\begin{eqnarray}
&& [N_{[p]}+ e^{\ell  p_0} N_{[p^\prime]}+ e^{\ell  (p_0+p^\prime_0)} N_{[p^{\prime \prime}]},
p_1 + e^{\ell p_0}p^\prime_1 + e^{\ell  (p_0+p^\prime_0)}p^{\prime \prime}_1]
 =\nonumber\\
&& ~~~~~~~~ = \left(\frac{e^{2 \ell p_0 } -1}{2\ell}  + \frac{\ell}{2} p_1^2 \right)
+ e^{2\ell p_0}
\left(\frac{e^{2 \ell p^{\prime}_0 } -1}{2\ell}  + \frac{\ell}{2} p^{\prime \, 2}_1 \right)
+e^{2\ell (p_0+p^{\prime}_0)}
\left(\frac{e^{2 \ell p^{\prime \prime}_0 }-1}{2\ell}  + \frac{\ell}{2} p^{\prime \prime \, 2}_1 \right)
+ \nonumber\\
&& ~~~~~~~~ ~~~~  + \ell e^{\ell p_0} p_1 p^\prime_1
+ \ell e^{\ell (p_0+p^{\prime}_0)} p_1 p^{\prime \prime}_1
+ \ell e^{\ell (2 p_0+ p^{\prime}_0)} p^\prime_1 p^{\prime \prime}_1 =
 \nonumber\\
&& ~~~~~ = \frac{\ell }{2}
 \left(p_1 + e^{\ell p_0} p^\prime_1 + e^{\ell  (p_0+p^\prime_0)}p^{\prime \prime}_1 \right)^2
  + \left(\frac{e^{2 \ell (p_0 + p^{\prime}_0 + p^{\prime \prime}_0) } -1}{2\ell}\right) = 0
\label{covathreeunoHHppkALL}
\end{eqnarray}
where for both these results I of course enforced on the right-hand side
the conservation law $p \oplus_\ell p^\prime \oplus_\ell p^{\prime \prime} = 0$ itself.

Note that from (\ref{covathreezeroHHppkALL}) and (\ref{covathreeunoHHppkALL})
it also follows that, for the conservation law $p \oplus_\ell p^\prime = 0$,
one has
$$[N_{[p]}+ e^{\ell  p_0} N_{[p^\prime]},
p_0 + p^\prime_0 ] = 0$$
and
$$ [N_{[p]}+ e^{\ell  p_0} N_{[p^\prime]},
p_1 + e^{\ell p_0}p^\prime_1 ] =0$$
So I have a fully consistent (and consistent to all orders in $\ell$)
DSR-relativistic description
of $p$-particles, propagating (with $p \oplus_\ell p^\prime = 0$ conservation)
and
interacting among themselves (with $p \oplus_\ell p^\prime  \oplus_\ell p^{\prime \prime} = 0$
conservation).

My next task is to introduce a second type of particles, with different DSR-relativistic
properties.
%I shall be satisfied to do that for a case where also for the second type of particle the
%deformation scale is $\ell$ but insisting that the DSR-relativistic properties be different.\\
For these ``$k$-particles" (whose momenta I shall consistently
denote with $k$ or $k^{\prime}$...)
I take the following on-shell relation
\begin{equation}
\cosh (\ell \mu) = \cosh (\ell k_0) - \frac{\ell^2}{2} k_1^2
\label{metrictorsyALLk}
\end{equation}
and boost such that
\begin{equation}
[N_{[k]}, k_0] =  k_1 ~,~~~
[N_{[k]}, k_1] = \frac{\sinh (\ell k_0)}{\ell}~,
\label{booststorsyALLk}
\end{equation}
which indeed is compatible with the on-shell relation
\begin{equation}
[N_{[k]},  \cosh (\ell k_0) - \frac{\ell^2}{2}  k_1^2 ]
 = 0 ~.
\label{invariantshelltorsyALLk}
\end{equation}
Notice that I am now considering a case where also the DSR-deformation of the second type
of particle is characterized by the same deformation scale $\ell$, but the form of
the  laws of transformation that apply to the two types of particles are very significantly different.

For this section of ``aside beyond leading order" I do not go as far as introducing
a proper composition law for $k$-particles and/or a proper ``$p$-particle/$k$-particle
mixing composition law"
to be used as general rules applicable to all sorts of processes. I will instead
directly show that some acceptable conservation laws, relevant for certain
 specific processes, do admit a fully relativistic
description.\\
And I will not elaborate on the ways to construct such conservation laws, but the careful
reader will notice that their structure can be seen as inspired by either one of two ways
of characterizing equivalently the covariance of my choice of
conservation law for ``$k$-particle propagation processes", which is
the undeformed one:
\begin{eqnarray}
k_0 + k^\prime_0 =0~,~~~k_1 + k^\prime_1 =0~.
\label{consKv1}
\end{eqnarray}
This evidently is compatible with a correspondingly undeformed law
of ``composition of boosts"  $N_{[k]}$:
\begin{eqnarray}
[N_{[k]}+N_{[k^\prime]}, k_0 + k^\prime_0]= k_1 + k^\prime_1 = 0 ~,~~~
[N_{[k]}+N_{[k^\prime]}, k_1 + k^\prime_1]
= \frac{\sinh (\ell k_0)}{\ell} + \frac{\sinh (\ell k^\prime_0)}{\ell} = 0 ~,
\label{consKv1BOOST}
\end{eqnarray}
where I of course also used the conservation law $k_0 + k^\prime_0 =0$, $k_1 + k^\prime_1 =0$ itself.

The anticipated interesting alternative way to characterize
this conservation law for ``$k$-particle propagation processes"
 is centered on noticing that (\ref{consKv1})
can be rewritten equivalently as
\begin{eqnarray}
k_0 + k^\prime_0 =0~,~~~e^{- \ell k^\prime_0 /2}k_1 + e^{\ell k_0 /2}k^\prime_1 =0~,
\label{consKv2}
\end{eqnarray}
whose compatibility with the $N_{[k]}$ boost could be described as follows
\begin{eqnarray}
&& [e^{- \ell k^\prime_0 /2} N_{[k]}+ e^{\ell k_0 /2}N_{[k^\prime]}, k_0 + k^\prime_0]=
e^{- \ell k^\prime_0 /2}k_1 + e^{\ell k_0 /2}k^\prime_1 =0~,\label{consKv2BOOSTa}\\
&&
[e^{- \ell k^\prime_0 /2} N_{[k]}+ e^{\ell k_0 /2}N_{[k^\prime]},
e^{- \ell k^\prime_0 /2}k_1 + e^{\ell k_0 /2}k^\prime_1 ]
= e^{- \ell k^\prime_0 /2} \frac{\sinh (\ell k_0)}{\ell}
+  e^{\ell k_0 /2} \frac{\sinh (\ell k^\prime_0)}{\ell} = 0
~.\label{consKv2BOOSTb}
\end{eqnarray}

I do not provide any other conservation law for processes involving
exclusively $k$-particles (as such I provide a picture for $k$-particles that do not
self-interact). But I do provide  ``mixing conservation laws", for
processes involving both $k$-particles and $p$-particles, with structure which may
be viewed as inspired either by the form of (\ref{consKv1}),(\ref{consKv1BOOST})
or by the form of (\ref{consKv2}),(\ref{consKv2BOOSTa}),(\ref{consKv2BOOSTb}).

The first such ``mixing composition law" which I exhibit
 is for simple ``oscillation processes"
$$k_0+p_0=0~,~~k_1+e^{\ell k_0/2}p_1=0$$
and is compatible with the following law of ``mixing of boosts"
$$N_{[k]}+ e^{\ell k_0/2} N_{[p]}$$
This is easily verified from the following two observations:
\begin{eqnarray}
&& [N_{[k]}+ e^{\ell  k_0/2} N_{[p]},
k_0+p_0]
= k_1 + e^{\ell k_0/2} p_1 = 0
\nonumber
\end{eqnarray}
and
\begin{eqnarray}
&& [N_{[k]}+ e^{\ell  k_0/2} N_{[p]} ,
k_1 + e^{\ell k_0/2} p_1]
 =\nonumber\\
&& ~~~~~~~~ = \frac{\sinh (\ell k_0)}{\ell}
+ e^{ \ell k_0 } \left(\frac{e^{2 \ell p_0 } -1}{2\ell}  + \frac{\ell}{2} p_1^2 \right)
+ \nonumber\\
&& ~~~~~~~~ ~~~~  + \frac{\ell}{2} e^{\ell k_0/2} k_1 p_1 =
 \nonumber\\
&& ~~~~~ = \frac{\ell }{2}
e^{\ell k_0/2} p_1
\left(k_1 + e^{\ell k_0/2} p_1 \right)
+\left(\frac{e^{\ell k_0 + 2 \ell p_0 } - e^{- \ell k_0}}{2\ell}\right) = 0
\nonumber
\end{eqnarray}

A first example of consistently relativistic conservation law for interactions
among $k$-particles and $p$-particles is the following:
$$k_0+p_0+p^\prime_0=0~,~~ k_1 + e^{\ell k_0/2} p_1 + e^{\ell k_0/2} e^{\ell p_0} p^\prime_1=0~,$$
which is for one $k$-particle interacting with two $p$-particles
and is compatible with the following law of ``kpp mixing of boosts"
$$N_{[k]}+ e^{\ell k_0/2} N_{[p]}+ e^{\ell k_0/2} e^{\ell p_0} N_{[p^\prime]}~.$$
This is easily verified as follows:
\begin{eqnarray}
&& [N_{[k]}+ e^{\ell  k_0/2} N_{[p]} + e^{\ell k_0/2} e^{\ell p_0} N_{[p^\prime]},
k_0+p_0+p^\prime_0]
= k_1 + e^{\ell k_0/2} p_1 + e^{\ell k_0/2} e^{\ell p_0} p^\prime_1 = 0
\nonumber
\end{eqnarray}
\begin{eqnarray}
&& [N_{[k]}+ e^{\ell  k_0/2} N_{[p]} + e^{\ell k_0/2} e^{\ell p_0} N_{[p^\prime]},
k_1 + e^{\ell k_0/2} p_1 + e^{\ell k_0/2} e^{\ell p_0} p^\prime_1]
 =\nonumber\\
&& ~~~~~~~~ = \frac{\sinh (\ell k_0)}{\ell}
+ e^{ \ell k_0 } \left(\frac{e^{2 \ell p_0 } -1}{2\ell}  + \frac{\ell}{2} p_1^2 \right)
+ e^{ \ell k_0 } e^{2\ell p_0}
\left(\frac{e^{2 \ell p^{\prime}_0 } -1}{2\ell}  + \frac{\ell}{2} p^{\prime \, 2}_1 \right)
+ \nonumber\\
&& ~~~~~~~~ ~~~~  + \frac{\ell}{2} e^{\ell k_0/2} k_1 p_1
+  \frac{\ell}{2} e^{\ell k_0/2} e^{\ell p_0} k_1 p^{\prime}_1
+ \ell e^{\ell k_0} e^{\ell p_0} p_1 p^{\prime}_1 =
 \nonumber\\
&& ~~~~~ = \frac{\ell }{2}
\left(e^{\ell k_0/2} p_1 + e^{\ell k_0/2} e^{\ell p_0} p^\prime_1 \right)
\left(k_1 + e^{\ell k_0/2} p_1 + e^{\ell k_0/2} e^{\ell p_0} p^\prime_1 \right)
+\left(\frac{e^{\ell k_0 + 2 \ell p_0 + 2 \ell p^{\prime}_0 } - e^{- \ell k_0}}{2\ell}\right) = 0
\nonumber
\end{eqnarray}

Finally I also exhibit a consistently relativistic conservation law for interactions
involving two $k$-particles and one $p$-particle:
$$k_0+p_0+p^\prime_0=0~,~~ e^{-\ell k^\prime_0} e^{-\ell k_0/2}k_1 +
 e^{-\ell k^\prime_0/2}k_1 + p_1=0$$
which is compatible with the following law of ``kkp mixing of boosts"
$$e^{-\ell k^\prime_0} e^{-\ell k_0/2} N_{[k]}+  e^{-\ell k^\prime_0/2} N_{[k^\prime]} +  N_{[p]}$$
as one can  again easily verify:
\begin{eqnarray}
&& [e^{-\ell k^\prime_0} e^{-\ell k_0/2} N_{[k]}+  e^{-\ell k^\prime_0/2} N_{[k^\prime]} +  N_{[p]},
k_0+k^\prime_0+p_0]
= e^{-\ell k^\prime_0} e^{-\ell k_0/2}k_1 +
 e^{-\ell k^\prime_0/2}k_1 + p_1= 0
\nonumber
\end{eqnarray}
\begin{eqnarray}
&& [e^{-\ell k^\prime_0} e^{-\ell k_0/2} N_{[k]}+  e^{-\ell k^\prime_0/2} N_{[k^\prime]} +  N_{[p]},
e^{-\ell k^\prime_0} e^{-\ell k_0/2}k_1 +
 e^{-\ell k^\prime_0/2}k_1 + p_1]
 =\nonumber\\
&& ~~~~~ = e^{- 2 \ell k^\prime_0} e^{-\ell k_0} \frac{\sinh (\ell k_0)}{\ell}+
e^{- \ell k^\prime_0}  \frac{\sinh (\ell k^\prime_0)}{\ell}
+ \left(\frac{e^{2 \ell p_0 } -1}{2\ell}  + \frac{\ell}{2} p_1^2 \right)
- \frac{\ell}{2} e^{- 2 \ell k^\prime_0} e^{-\ell k_0} k_1^2
- \frac{\ell}{2} e^{- \ell k^\prime_0} k_1^{\prime \, 2}
- \ell e^{- 3 \ell k^\prime_0/2} e^{-\ell k_0} k_1 k_1^{\prime}
= ~~~~~~~~~ \nonumber\\
&& ~~~~~ = \frac{\ell }{2}
\left(p_1 - e^{- \ell k^\prime_0} e^{-\ell k_0/2} k_1 -
e^{- \ell k^\prime_0/2} k_1^{\prime} \right)
\left(p_1 + e^{- \ell k^\prime_0} e^{-\ell k_0/2} k_1 +
e^{- \ell k^\prime_0/2} k_1^{\prime} \right)
+\left(\frac{e^{2 \ell k_0 + 2 \ell p_0 } - e^{- 2 \ell k^\prime_0}}{2\ell}\right) = 0
\nonumber
\end{eqnarray}

\section{Closing remarks}\label{closingsec}
The fact, here established, that there are consistent scenarios of ``non-universal"
deformation of Lorentz symmetry,
with particle-type-dependent deformations of relativistic kinematics,
can be used to address some long-standing
issues in DSR research, mentioned in Sec.~I, and can also be the starting point
for several further developments and further generalizations.

Some of these future studies could be directed toward the understanding of
the associated relativity of spacetime locality. It
is established that already for some ``universal" DSR-deformation
schemes spacetime locality becomes relative~\cite{bob,leeINERTIALlimit,k-bob}.
The case of ``non-universal" DSR-deformation here introduced should have subtle implications
also for the characterization of the relativity of spacetime locality.\\
And it is established that for some schemes of ``universal" DSR deformation
it is possible to provide a formulation within the ``relative-locality framework"~\cite{prl,grf2nd},
centered on the geometry of momentum space. It would therefore be interesting to attempt to generalize
the relative-locality-framework formulation also to the case
of the ``non-universal" DSR deformations here introduced, but this raises some intriguing questions:
could then one describe the different types of particles on the same geometry of momentum space?
or should one rather seek a formulation based on different momentum-space geometries for
different particles? and in that case which sort of geometric requirement could codify
the feature here formulated in terms of a ``mixing composition law"?

One more intriguing question I want to mention
 here
concerns  the types of different relativistic properties which can be found
to be compatible.
In the construction of the cases I here analyzed as illustrative examples
a consistent particle-dependent (``non-universal") DSR deformation
was achieved also exploiting in part the fact that I confined myself to considering
only rather mild differences of relativistic
properties: I managed to ``mix" particles governed by quantitatively very different
deformations of  relativistic
kinematics, but all members of a class of related such deformations.
Having established that this can be done consistently, one may now ask whether it is
possible to ``mix" in a single consistent relativistic framework particles with more profoundly
different relativistic properties. Ideally one would like to establish some sort
of theorem characterizing the types of different relativistic properties
which can be made compatible in the sense I here introduced. Such a theorem
appears to be extremely challenging, but even gaining some expertise on the basis
of a few ``trial-and-error exercises" might be valuable.

\end{document}